\documentclass{sigchi}

\toappear{\scriptsize Permission to make digital or hard copies of all or part of this work for personal or classroom use is granted without fee provided that copies are not made or distributed for profit or commercial advantage and that copies bear this notice and the full citation on the first page. Copyrights for components of this work owned by others than ACM must be honored. Abstracting with credit is permitted. To copy otherwise, or republish, to post on servers or to redistribute to lists, requires prior specific permission and/or a fee. Request permissions from permissions@acm.org. \\
{\emph{CHI 2016}}, May 7--12, 2016, San Jose, California, USA. \\
Copyright \copyright~2016 ACM ISBN 978-1-4503-3362-7/16/05\ ...\$15.00. \\
http://dx.doi.org/10.1145/2739XXX.XXXXX}
\usepackage{balance}  % to better equalize the last page
\usepackage{graphics} % for EPS, load graphicx instead 
\usepackage{txfonts}
\usepackage{times}    % comment if you want LaTeX's default font
\PassOptionsToPackage{hyphens}{url}      % llt: nicely formatted URLs
\usepackage[pdftex]{hyperref}
\usepackage{color}
\usepackage{textcomp}
\usepackage{booktabs}
\usepackage{ccicons}
\usepackage{todonotes}
\usepackage{framed}
\usepackage{tikz}
\usepackage{varwidth}
\usepackage{multirow}
\usepackage{amsmath}
\makeatletter
\def\url@leostyle{%
  \@ifundefined{selectfont}{\def\UrlFont{\sf}}{\def\UrlFont{\bf\ttfamily}}}
\makeatother
\def\pprw{8.5in}
\def\pprh{11in}

\definecolor{linkColor}{RGB}{6,125,233}
\hypersetup{%
  pdftitle={SIGCHI Conference Proceedings Format},
  pdfauthor={LaTeX},
  pdfkeywords={SIGCHI, proceedings, archival format},
  bookmarksnumbered,
  pdfstartview={FitH},
  colorlinks,
  citecolor=black,
  filecolor=black,
  linkcolor=black,
  urlcolor=linkColor,
  breaklinks=true,
}

\begin{document}

\title{The Proficiency-Congruency Dilemma: Virtual Team\\Design and Performance in Multiplayer Online Games}

\numberofauthors{1}
\author{
	\alignauthor Jooyeon Kim,$^{1}$ Brian C. Keegan,$^{2}$ Sungjoon Park,$^{1}$ Alice Oh $^{1}$\\
     \affaddr{$^{1}$ School of Computing, KAIST, Daejeon, Republic of Korea}\\
     \affaddr{$^{2}$ Harvard Business School, Boston, Massachusetts, USA}\\
     \email{jooyeon.kim@kaist.ac.kr}, \email{bkeegan@acm.org}, \email{sungjoon.park@kaist.ac.kr}, \email{alice.oh@kaist.edu}
}

%\numberofauthors{3}
%\author{%
%  \alignauthor{Jooyeon Kim\\
%    \affaddr{Affiliation}\\
%    \affaddr{City, Country}\\
%    \email{e-mail address}}\\
%  \alignauthor{2nd Author Name\\
%    \affaddr{Affiliation}\\
%    \affaddr{City, Country}\\
%    \email{e-mail address}}\\
%  \alignauthor{3rd Author Name\\
%    \affaddr{Affiliation}\\
%    \affaddr{City, Country}\\
%    \email{e-mail address}}\\
%}

\maketitle

\begin{abstract}
Multiplayer online battle arena games provide an excellent opportunity to study team performance. When designing a team, players must negotiate a \textit{proficiency-congruency dilemma} between selecting roles that best match their experience and roles that best complement the existing roles on the team. We adopt a mixed-methods approach to explore how users negotiate this dilemma. Using data from  \textit{League of Legends}, we define a similarity space to operationalize team design constructs about role proficiency, generality, and congruency. We collect publicly available data from 3.36 million users to test the influence of these constructs on team performance. We also conduct focus groups with novice and elite players to understand how players' team design practices vary with expertise. We find that player proficiency increases team performance more than team congruency. These findings have implications for players, designers, and theorists about how to recommend team designs that jointly prioritize individuals' expertise and teams' compatibility.
\end{abstract}


%\category{K.4.3}{Computers and Society}{Organizational Impacts ()}
%\category{J.4}{Computer Applications}{Social and Behavioral Sciences (Psychology; Sociology)}

%%%%%%%%%%%%%%%%%
% Introduction %
%%%%%%%%%%%%%%%%%

\section{Introduction}

How does team composition influence team performance? Organizational researchers have explored this question for decades, generating a rich body of literature around themes such as diversity, cohesion, deviance, and assembly mechanisms. Sports traditionally provided an excellent empirical setting to examine organizational behavior in teams owing to the consistency of rules, clarity of outcomes, repeated observations, and behaviors at the levels of individuals and groups~\cite{day_sporting_2012, frey_sport_1991, wolfe_sport_2005}. But the sports metaphor falls apart when we want to understand team performance in the context of virtual work, temporary groups, and self-assembled teams using information technology to support organizational processes.% that increasingly characterize our globalized network society~\cite{benkler_wealth_2006, castells_rise_2011}. 

We argue an emerging video game genre called multiplayer online battle arenas (MOBAs) have substantial potential to bridge the gap between traditional sports metaphors and contemporary networked organizations by providing new insights into organizational behavior by analyzing the large-scale and fine-grained digital traces of in-game player behavior~\cite{williams_mapping_2010}. While emerging game research emphasizes the importance of ``communication'' and ``teamwork'' as determinants of team success in MOBAs~\cite{blackburn_stfu_2014, iosup_implicit_2014, johnson_all_2015, kwak_cyberbullying_2015, shores_identification_2014}, these constructs are under-developed despite the extensive literature on their role in team performance~\cite{cohen_makes_1997, guzzo_teams_1996, kozlowski_work_2003, kozlowski_enhancing_2006, mathieu_team_2008}. Drawing from the organizational behavior literature, we define several quantitative constructs to test research questions about self-assembled team design processes.

 %In \textit{LoL}, two five-player teams compete to destroy against each other's base while defending their own over the course of an approximately thirty minute match. Each player selects a single ``champion'' from a cast of 126\footnote{As of September 2015. \url{http://gameinfo.na.leagueoflegends.com/en/game-info/champions/}} to play for the entire match in order to fulfill specific offensive, defensive, and supportive functional roles. 

We test the relationship between these constructs and group performance in the most popular MOBA game in the world, \textit{League of Legends} (\textit{LoL}). In order to be successful, players must negotiate a \textit{proficiency-congruency dilemma} when selecting the champions to assemble their five player teams. Analogous to team-sport athletes who often possess specific strengths and thus excel at certain positions, \textit{LoL} users control ``champions'' that differ in their skill sets and the roles that they can best fulfill. On one hand, players can select champions that they are more skilled at or enjoy playing, even if this duplicates other roles already represented on the team. On the other hand, players can select champions that are most congruent with others' abilities, even if they personally lack experience playing those champions or in that role. 

This basic social dilemma motivates the research questions for this paper: Do teams following proficiency-maximizing strategies perform better or worse than teams employing congruency-maximizing strategies? How do elite players negotiate this dilemma compared to average players? How do players talk about this dilemma when assembling their teams? To explore these research questions, we employed a mixture of methods using data mining, surveys, machine learning, and participant observations and interviews. We collected publicly-accessible data from 1.94 million \textit{LoL} matches containing 3.36 million users from across North America, Europe, and Korea, surveyed 492 players to validate key constructs, and observed and interviewed 10 expert and 10 novice users as they played the game. 
 
%%%%%%%%%%%%%%
% Background %
%%%%%%%%%%%%%%

\section{Background}
%In this section, we briefly describe LoL and we highlight the team performance literature.

%The importance of teams can be found throughout culture, politics, economics, and society, which has motivated research generating thousands of studies across the social and---increasingly---the information sciences into understanding team performance and effectiveness. Reviews of the literature on team effectiveness emphasize a variety of cognitive, motivational/affective, and behavioral processes that allow team members to come together to combine resources, resolve coordination demands, and execute their tasks~\cite{kozlowski_enhancing_2006}.  
% As work becomes more distributed, temporary, and generally complex, traditional individualized roles are being replaced by teams that provide more diverse skills, greater expertise, and quicker responses.

%Teams are defined as two or more individuals who (1) socially interact, (2) pursue of common goals, (3) perform tasks for an organization, (4) share workflow and outcome interdependencies, (5) have distinct roles and responsibilities, and (5) are embedded within a larger organizational system~\cite{cohen_makes_1997}. Unlike production or service teams, project and action teams that operate within changing environments, which impose task demands, which require coordination, cooperation, and communication processes to manage, all of which influence effectiveness over a cycle of task preparation, engagement, and disengagement~\cite{kozlowski_work_2003, kozlowski_enhancing_2006, mathieu_team_2008, sundstrom_work_2000}. 

Many sports and games are defined by their need for teamwork, which has made them a natural setting for empirical research about team performance. While many organizations inherently display ``messiness'' in their rules, roles, outcomes, and other important factors, sports teams have precisely-defined rules, stable roles, clear outcomes, strong incentives, distinct boundaries, shared identity, and meticulous record-keeping that make them a  ``model organism'' for studying team behavior~\cite{day_sporting_2012, frey_sport_1991, wolfe_sport_2005}. However, many of the features that make sports a unique empirical model bear less-and-less resemblance to contemporary organizations where teams are increasingly distributed, virtual, self-assembled, cross-functional, and fluid/temporary. Team-based multiplayer online games offer many of the same benefits as sports teams for research about team performance without some of those limitations on generalizability~\cite{yee_labor_2006}, making them potentially more ideal ``model organisms'' for team effectiveness research in the ``wild'' outside of laboratories~\cite{elnasr_game_2013, williams_mapping_2010}.

To better understand the potential of e-sports games to be persuasive models of team effectiveness, we review the team effectiveness research with a particular focus on team design and assembly mechanisms in the context of team member expertise and skill as well as team diversity and cohesion. This literature is largely drawn from social psychology and organizational behavior, but we contextualize it with emerging research from virtual and multiplayer game teams such as multiplayer online battle arenas (MOBAs). Because the teams assembled in online games are more dynamic and mutable than in traditional organizations, this context provides a boundary case to examine a fundamental but overlooked trade-off that we call the \textit{proficiency-congruency dilemma}. 

%Based on this dilemma, we identify a set of general hypotheses and contextualize them within the specific genre of competitive online ``e-sports'' games called .

% Much of this literature emphasizes the importance of differentiating (rather than aggregating) individual- and team-level constructs because team members pursue a combination of goals~\cite{deshon_multiplegoal_2004}, but it also often assumes constructs such as skill or diversity are fixed and unchanging. 

\subsection{Team design and assembly}
Team design addresses the issues of how to (1) assemble teams from the available human and material resources in an organization, (2) set team goals based on the organizational context, and (3) incentivize teams to most effectively perform their tasks~\cite{stewart_metaanalytic_2006}. Teams must be designed in response to the complexity of the work, discretion over how to accomplish the work, and uncertainty in the team's environment~\cite{goodman_familiarity_1992}. Other factors such as leadership, mutual monitoring, backup behavior, adaptability, trust, and heedfulness are also important for performance~\cite{salas_is_2005}.

Team members have both individual- and team-level goals they will pursue. If the incentives reward individual rather than team performance, then team performance will likely suffer. Conversely, teams that have acquired stronger self-regulation strategies to align individual and team-level interests will out-perform teams that do not align goals across levels~\cite{deshon_multiplegoal_2004}. There are multiple methods for aligning team incentives such as maximization, balance, redundancy, or complementarity~\cite{edwards_personenvironment_2008, muchinsky_what_1987, seong_person_2012} but their effects on team performance depends on task and context~\cite{harrison_time_2003, kozlowski_enhancing_2006}. 

%MOBAs like \textit{LoL} to exhibit high levels of complexity given the role differentiation and expertise required, high levels of discretion in terms of how to execute the tasks, but low levels of uncertainty since the opponents' team composition and skill is observable. 

MOBAs like \textit{LoL} employ a number of cooperative game design patterns that require players to employ complementary characters, shared goals, and synergistic abilities~\cite{seif_el-nasr_understanding_2010}. Like other temporary teams that coordinate by regenerating prior role structures~\cite{bardzell_blissfully_2008, bechky_gaffers_2006, valentine_team_2014}, MOBA players typically follow a consistent strategy from game-to-game that prescribes specific team compositions and champion assignments to roles~\cite{donaldson_mechanics_2015, kou_playing_2014}. This ``meta'' strategy makes many of the later stages in classic team development models (e.g., ``forming, storming, norming, performing''~\cite{tuckman_stages_1977}) entirely path dependent upon the initial formation stage since champions' roles are fixed and players cannot change champions mid-game. 

\subsection{Skills and expertise as proficiency}

A fundamental component of team performance is the skill and expertise of its constituent team members. Expertise is defined as the ability to consistently perform a set of domain-relevant tasks. Experts across domains exhibit similar characteristics such as substantial domain knowledge, enhanced pattern recognition/sensemaking, rapid task execution ability, improved recall, novel knowledge representations, and high levels of self-monitoring but expertise also comes with cost such as requiring a substantial investment of time and difficulty transferring skills to other domains~\cite{ericsson_expert_1996, chi_overview_1988}. 

Teams members that are familiar with specific skills, tasks, and roles should perform better than teams lacking this familiarity and expertise~\cite{goodman_familiarity_1992, huckman_team_2008}. Players may develop these different skills in response to self-efficacy feedback loops operating at either the individual or team levels. Self-efficacy is a process by which individuals or teams evaluate their performance by comparing their progress and goals~\cite{bandura_selfefficacy_1977}. In the absence of traditional support mechanisms, individual self-efficacy is especially important for virtual teams~\cite{johnson_realities_2009, staples_exploring_2007}. 

The development of expertise in video games follows similar patterns as other domains such as chess emphasizing the sustained and intensive practice that cause psychological, cognitive, and physiological changes~\cite{ericsson_expert_1996, latham_just_2013}. Expert players in multiplayer games distinguish themselves from novice players by employing distinctive communication strategies and play styles~\cite{huffaker_social_2009, reeves_experts_2009}. Because MOBA players are often members of completely different teams across matches, players are incentivized to develop \textit{individual} self-efficacy, such as developing champion-specific tactics, rather than to develop \textit{team} self-efficacy, such as general tactics like minion-farming~\cite{kou_playing_2014}. The steep learning curve and high degree of difficulty to master the tactical mechanics of any single champion are major motivators for players~\cite{johnson_all_2015}.

% A crucial part of self efficacy is \textit{enactive mastery} whereby successful experiences become indicators of capability and help people acquire new skills more than failure experiences~\cite{bandura_selfefficacy_1997}. 

%This has crucial implications since team efficacy has a stronger relationship with team performance on tasks with high levels of interdependence, such as integrating strategies and sharing effort, compared to tasks with low interdependence, such as simply pooling contributions~\cite{gully_metaanalysis_2002}. 

%\vspace{.5em}
%\begin{framed}
%\textbf{H1a}: Non-elite players develop proficiency-maximizing skills more than diversity-maximizing skills.
%\end{framed}

\vspace{.5em}
\begin{framed}
\textbf{H1}: Teams with members playing roles similar to their previous roles will perform better than teams with members playing roles different from their previous roles.
\end{framed}

\subsection{Diversity and cohesion as congruency}

Individual team members' knowledge, skills, and abilities influence their ability to coordinate and problem solve with team members~\cite{stevens_knowledge_1994}. However, ``taskwork'', or \textit{what} teams do, is distinct from ``teamwork'', or \textit{how} they do it with each other~\cite{marks_temporally_2001} and both have positive effects on team performance and effectiveness~\cite{cohen_makes_1997, kozlowski_enhancing_2006, mathieu_team_2008}. Team cohesion reflects the interpersonal bonds, commitment, and shared identity~\cite{mudrack_defining_1989} that emerge from agreeableness and communication~\cite{bradley_team_2013}.

%A self-evident but frequently overlooked measure of team performance is the process of ensuring the persistence and cohesion of the team in the face of conflict and challenges.

However, teams composed of unfamiliar members or facing fluid tasks may inhibit team performance because of new complexity, altered information flows, and disrupted pacing unless team members have diverse experience that complements and/or supplements each other~\cite{huckman_fluid_2011, huckman_team_2008, muchinsky_what_1987}. While diversity of roles can improve a team's information processing capacity and broaden its knowledge, skills, and abilities, these depend on the relative breadth and depth of team members' individual experience and expertise~\cite{horwitz_effects_2007}. Teams having specialists focused on task execution and teams having generalists drawing on a variety of experiences will vary in their effectiveness depending on the work context~\cite{bunderson_comparing_2002, derue_search_2007, kristof-brown_consequences_2005}.

Group cohesion, even in temporary teams in online games, also affects performance~\cite{huang_mastering_2013}. Cooperation within virtual worlds and online games unfolds through shared attentiveness, responsiveness, functional identities, focus, and objectives to action within a shared visual space~\cite{williams_coordinated_2012}. The challenges and satisfaction of teamwork in MOBAs is a recurring finding as users enjoy playing with friends, performing well as a team, and varying their play style based on the context~\cite{eggert_classification_2015, iosup_implicit_2014, johnson_all_2015, ong2015player, shores_identification_2014}. Despite the temporary nature of the team, MOBA players take actions to create a collaborative environment, share leadership, and hold deviant players accountable~\cite{kou_playing_2014}. High-performing teams in MOBAs also show an increased spatio-temporal mobility and more rapid ability to converge, suggesting differences in situational awareness and disposition to cooperate~\cite{drachen_skill-based_2014}.

\vspace{.5em}
\begin{framed}
\textbf{H2}: Teams with members playing roles congruent with other team members' roles will perform better than teams containing less congruent roles.
\end{framed}

\subsection{Proficiency-Congruency Dilemma}
%\begin{figure}[t]
%\centering
%\begin{tikzpicture}[
%        scale=1.5,
%        dot/.style={circle,
%                    minimum size=8pt,inner sep=0pt,
%                    outer sep=-1pt}]
%    % Draw axes
%    \draw [ultra thick,->] (0,0) -- (0,1) node [above] {\begin{varwidth}{1.5in}\centering General knowledge,\\skills, abilities\end{varwidth}};
%    \draw [ultra thick,->] (0,0) -- (0,-1) node [below] {\begin{varwidth}{1.5in}\centering Specific knowledge,\\skills, abilities\end{varwidth}};
%    \draw [ultra thick,->] (0,0) -- (1,0) node [right] {\begin{varwidth}{1in}\centering Team\\processes\end{varwidth}};
%    \draw [ultra thick,->] (0,0) -- (-1,0) node [left] {\begin{varwidth}{1in}\centering Task\\processes\end{varwidth}};
%    % Draw dots
%    \node[dot,fill=blue,{label=[blue]below:Congruency}] at (.8,.66) {};
%    \node[dot,fill=red,{label=[red]above:Proficiency}] at (-.8,-.66) {};
%\end{tikzpicture}
%\caption{Team competency framework. Congruency-maximizing strategies are shown in blue in the upper-right quadrant, proficiency-maximizing strategies are shown in red in the lower-left quadrant.}
%\label{fig:two_by_two}
%\end{figure}

High-performing team must balance many trade-offs involving expertise and diversity to create a team that ``fits'' together. Person-environment fit, where an individual and task environment are well-matched, is distinct from person-group fit, where an individual is compatible with a team~\cite{edwards_personenvironment_2008, kristof_person-organization_1996, kristof-brown_consequences_2005}. Compatibility can reflect \textit{supplementary} fits, where a person matches or reinforces expertise, attitudes, and roles that already exist on a team, or \textit{complementary} fits where a member offsets or fills in gaps or weaknesses on the team~\cite{muchinsky_what_1987, seong_person_2012}. 

The \textit{proficiency-congruency dilemma} is an optimization challenge faced during team design and assembly about how to make members' expertise and diversity fit together most effectively. Teams can pursue their comparative advantage by employing members whose competencies \textit{supplement} existing strengths or attempt to reduce risk through diversification by employing members whose competencies \textit{complement} current weaknesses~\cite{cannon_defining_1995}. As in other social dilemmas, individuals have incentives to prioritize their individual self-interest (\textit{e.g.}, playing roles maximizing their proficiency) at the cost of collective benefits (\textit{e.g.}, selecting incongruent roles that decrease team performance)~\cite{kollock_social_1998}. %The first dimension captures how team members balance between focusing on ``task work'' processes versus ``team work'' processes. The second dimension captures how team members must balance between developing knowledge, skills, and abilities that are specific or general. Figure~\ref{fig:two_by_two} illustrates the proficiency-congruency dilemma using this framework. 

In MOBAs like \textit{League of Legends}, team members developing skills specific to individual tasks, such as perfecting an action sequence combination for a single champion, are examples of \textit{proficiency}-maximizing strategies that neither generalize nor support team processes. Team members developing general skills that support team-level processes, such as learning the tactics to be effective in a support role, are examples of \textit{congruency}-maximizing strategies that neither demand attention to individual tasks nor specific expertise. A focus on proficiency skills may leave players ill-equipped to play other roles if their champion is unavailable while a focus on congruency skills may prevent players from fully maximizing the potential of a single champion.

\vspace{.5em}
\begin{framed}
\textbf{H3}: The team design choices of elite groups are better optimized for the proficiency-congruency dilemma than novice groups' choices.
\end{framed}

%%%%%%%%%%%%%%%%%
% Research Approach %
%%%%%%%%%%%%%%%%%

\section{Research Approach}

%\subsection{Research setting}

%\begin{figure}[t]
%\centering
%  \includegraphics[width=0.6\columnwidth]{figures/map.pdf}
%  \caption{Overview of LoL's classic battle arena, the Summoner's Rift. The image is provided by Riot API.}~\label{fig:map}
%\end{figure}

For this study, we use a set of game log data from the popular online video game, \textit{League of Legends} (\textit{LoL}). \textit{LoL} is published by Riot Games and is the most popular MOBA and the most-played PC game in the world, with more than 27 million daily users in 2014~\cite{riot_new_heights}. The classic battle arena\footnote{\textit{LoL} offers several maps, but the 5-vs.-5 ``Summoner's Rift'' map is the most popular and most representative of the MOBA genre.} in \textit{LoL} is a square map with three primary ``lanes'' (top, middle, and bottom) connected by a middle ``jungle'' area containing special bonuses as well as opportunities to ambush opponents. Individual players control a single unit and these ``champions'' vary substantially in abilities. Each team of five selects different champions for specific complementary offensive, defensive, and support roles and takes initial positions across the three lanes or jungle. All champions are initially weak but they accumulate wealth and experience by killing other champions, defensive structures, and waves of non-player ``minion'' characters. Players use this wealth and experience to purchase items and to unlock more powerful abilities to give them an advantage with the goal of destroying the opponent's base. A match takes 35 minutes on average to complete~\cite{harold_match}.

%\textit{LoL} has also become the most popular ``e-sport'' game: its 2014 World Championship attracted more than 32 million online users and \$2.3 million in prize money.~\cite{segal_lol_2014}

In \textit{LoL}, users can play in a league system that matches users with similar skill levels into the same group~\cite{wikia_league_system}. For each one-year season, users are assigned to a league containing several hundred users. Each league is broken into one of five tiers (in increasing order of level): Bronze (B), Silver (S), Gold (G), Platinum (P), and Diamond (D). Each league includes five subgroups, call divisions, that range from $I$ (high) to $V$ (low). Moreover, there are few hundred exceptional users who are assigned leagues in Master (M) and Challenger (C) tiers, which has only one division each~\cite{ranked_play_faq, wikia_league_system}. Users' seasonal win rates converge to 0.5; highly successful users in each tier and degree will earn a promotion to higher ranked groups, whereas unsuccessful ones will be demoted in the division, until their win rate stabilizes at around 0.5~\cite{riot_matchmaking}. Therefore, users' tier and division is regarded as an indicator of their long-term success; users will be promoted or demoted into higher/lower tiers and divisions until their win rate stabilizes at 0.5. The players' tiers are  Bronze (B), Silver (S), Gold (G), Platinum (P), and Diamond (D) levels.

%\begin{figure}[t]
%\centering
%\includegraphics[width=\columnwidth]{figures/maxresdefault.jpg}
%\caption{The Champion Selection screen.}\label{fig:champion_select}
%\end{figure}

A ranked game differs from the ``normal (unranked)'' play modes in \textit{LoL} because it follows a draft pick methodology, which has several consequences for how players select champions. Each player has a hidden matchmaking rating (MMR) that will ``attempt to assemble two teams as fairly as possible so both teams have an equal chance of winning''~\cite{ranked_play_faq}. Players' ``pick order'' within a team is randomized at the start of each match's selection process. In the first phase of the ranked champion selection, the first pick on each team bans three champions from either team playing. In the second phase, members of each team select champions alternating between teams in their respective pick order. An important factor here is that each champion can only be played by one team. In the third phase, when all members of both teams have picked their champions, team members can trade the selected champions so that the valuable early picks can be re-allocated to users most qualified to play them. %An example of the Champion Selection screen in Figure~\ref{fig:champion_select}. %The character selection method in the ranked game thus imposes some restrictions on the players' choice of the champions.

%A player can enjoy Summoner's Rift by three ways; normal, solo rank, and team rank mode. In normal mode, win-lose does not affect a player's ranking. But in solo rank mode, you are in a team with strangers and win-lose does affect your solo-rank tiers. Lastly, in team rank mode, you can play with your friends and win-lose does affect your team-rank tiers, which is independent to solo rank tiers. You are are always automatically matched to a similar level of opponent team.

%%%%%%%%%%%%%%%%%
% Experimental Setup %
%%%%%%%%%%%%%%%%%

%\section{Experimental Setup}
%We describe the data collecting method and basic statistics of our dataset, and discuss how the league of legends data is suited for analyzing virtual team studies. Further, we introduce our methodology of representing each champion in the game as a vector, and measuring distance between the champions. Finally, we show the validity of our methodology and argue that our computational approach of measuring champion distance is a reasonable and efficient alternative for large-scale user surveys.

\subsection{Data Collection}
We collected publicly-accessible player data using Riot API~\cite{riot_api_general}, which provides detailed information about players' match histories. For each tier, we extracted a random snowball-sample of players as seed players based on the matches they played, as well as summary statistics about their actions in those matches, such as champion selections, selection orders, kill-death-assist performance, and win-loss outcomes. We note that we can only identify the current tier and division information for the seed players. Because we are not able to track their previous records of tier and division, we gathered data on only their latest 60 matches. For the remaining players for whom we do not have the exact tier and division information, we assumed they belong to the same tier and division as the seed players with whom they played in the same matches. We gathered the data across three server regions (North America, Europe, and Korea) and made sure that the user size is evenly distributed throughout all tiers and divisions. The detailed numbers of players and matches are summarized in Table~\ref{tbl:regional_stats}. 

%and identified their league names. Then, we retrieved hundreds of \texttt{userids} within each league that functions as seed users in our dataset. Using the seed users, we retrieved logs of their recent matches which provide participant \texttt{userids}, their champion selections, selecting orders, and win-lose information. 

This dataset provides favorable experimental conditions for researchers to analyze both short-term virtual team performance and long-term individual success. First, the teams we examine in \textit{LoL} are algorithmically rather than socially formed: users' preferences are strictly excluded resulting in short-term, relation-free, and task-oriented teams. Second, as users are constantly re-assigned to new teams and opponents on the basis of skill similarity to create fair matches, skill biases are largely already controlled for and should not predict the short-term team performance. %Long-term individual success can be analyzed since individuals are formally represented by the tier and division information with 27 fine-grained separations.
%unlike other on-line groups that encompass both social interaction and task solving motivations~\cite{huang_functional_2013}; in LoL, users are randomly assigned and matched with strangers, and there is no guarantee that they will meet each other again. 

%\begin{figure}[t]
%\centering
%  \includegraphics[width=0.95\columnwidth]{figures/datasize_tier.pdf}
%  \caption{Data statistics by tiers. Users may appear in multiple tier groups.}~\label{fig:tier_stats}
%\end{figure}

\begin{table}[t]
\centering
\resizebox{0.45 \textwidth}{!}{%
\begin{tabular}{ c |  r r r r } 
            & \multicolumn{1}{c}{Korea} & \multicolumn{1}{c}{N. America} & \multicolumn{1}{c}{W. Europe} & \multicolumn{1}{c}{Total} \\ \hline
Seed players & 137                    & 129                    & 144                     & 410                       \\
Players      & 1,295,568              & 889,541                & 1,174,426               & 3,359,535                 \\
Matches       & 618,615                & 605,056                & 714,578                 & 1,938,249                \\ 
\end{tabular}}
\caption{Sample statistics by region.}
\label{tbl:regional_stats}
\end{table}

\subsection{Champion similarity space}
\textit{LoL} players select a single champion to play from a cast of more than 120 that vary substantially in abilities and role. These champions are designed so they can be a combination of play styles and in-game roles such as ``assassin'', ``fighter'', ``mage'', ``support'', ``tank'', and ``marksman''~\cite{riot_champions}. In order to develop constructs around players' proficiency and team congruency, we define and employ a \textit{champion similarity space} to measure the ``distance'' between champions in two different ways: feature distance and subjective distance.
%We use the feature distance for the remainder of the analyses rather than the subjective distance because it is less subject to sampling biases.

\subsubsection{Feature distance}
\begin{figure}[t!]
  \centering
  \includegraphics[width=0.6\columnwidth]{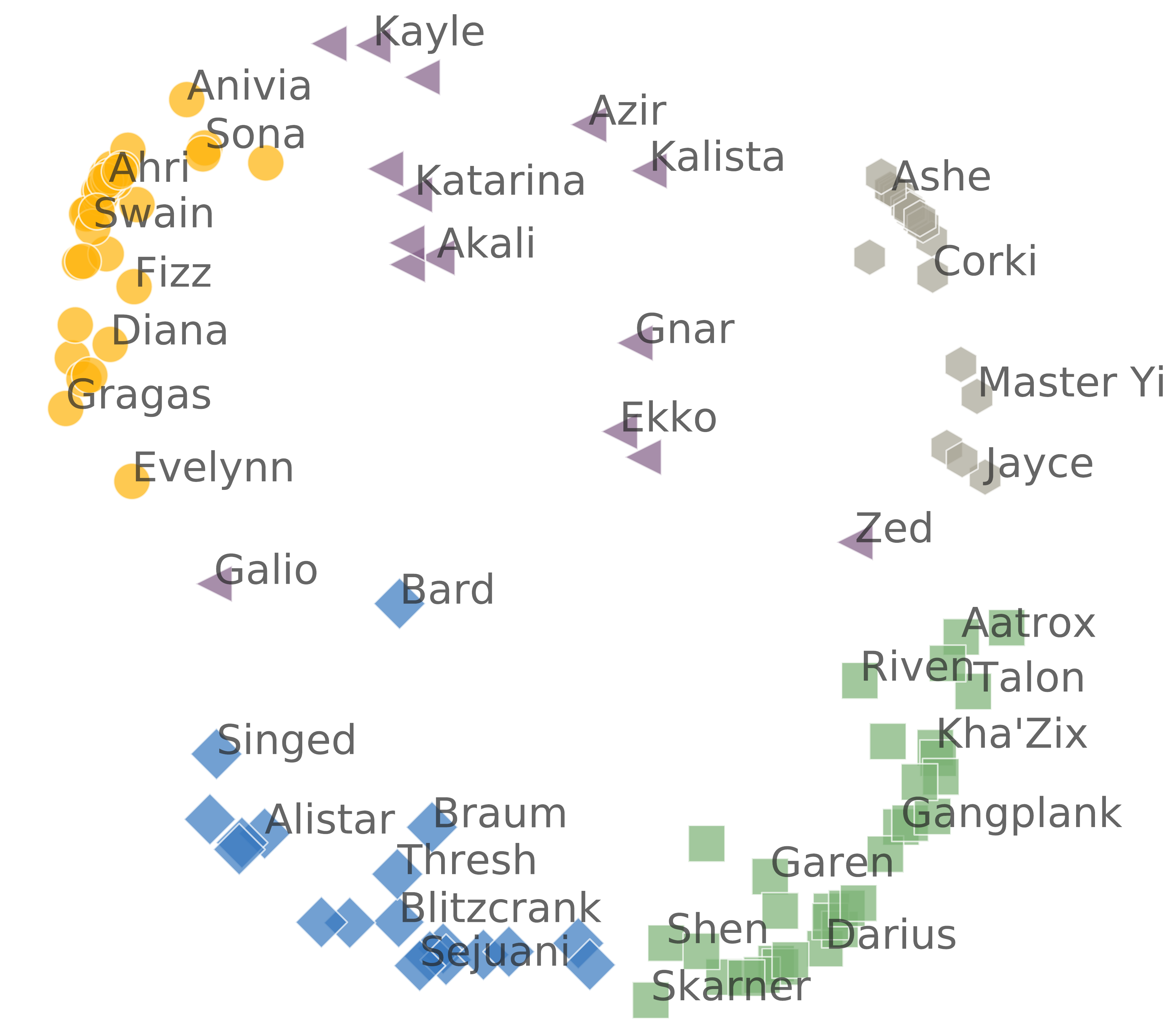}
  \caption{Champion functional clusters. We first used PCA to reduce the champion feature space to the most salient features, and then k-means to partition the champions into five clusters. We used multidimensional scaling to visualize the champions in the 2-dimensional space.}~\label{fig:champ_distance}
\end{figure}

Champions are defined by more than 174 features such as skill range, attack damage, and movement speed~\cite{riot_champions_data}~\footnote{Available at http://uilab.kaist.ac.kr/research/CHI2016}. We use principal components analysis (PCA) over this feature space and choose the first ten principal components.\footnote{We tried various numbers of principal components, and for all settings, the experimental results were consistent.} We measured the distance between every pair of champions using cosine distance of the ten principal components. We then apply multidimensional scaling to project the results into two-dimension and k-means clustering to partition the champions into five groups that align with the five spots on a team. Figure \ref{fig:champ_distance} illustrates these clusters.

\subsubsection{Subjective distance}
We surveyed LoL players to judge the similarity of champion pairs. In addition to validating the sensibility of our champion similarity metric based on feature distances, players' similarity judgements reflect their level of understanding about the LoL champions. We recruited 492 game players from the Bronze through Diamond tiers via an online survey, and they were asked to rate the similarity between two randomly chosen champions on a scale from 1 to 7. Players rated an average of 45 champion pairs for a total of 22,347 champion distance ratings. We discarded champion pairs with fewer than three user ratings and also ratings deviated by more than $\pm 2$ standard deviations from the mean value. 

\begin{figure}[t!]
 \centering
   \includegraphics[width=0.5\columnwidth]{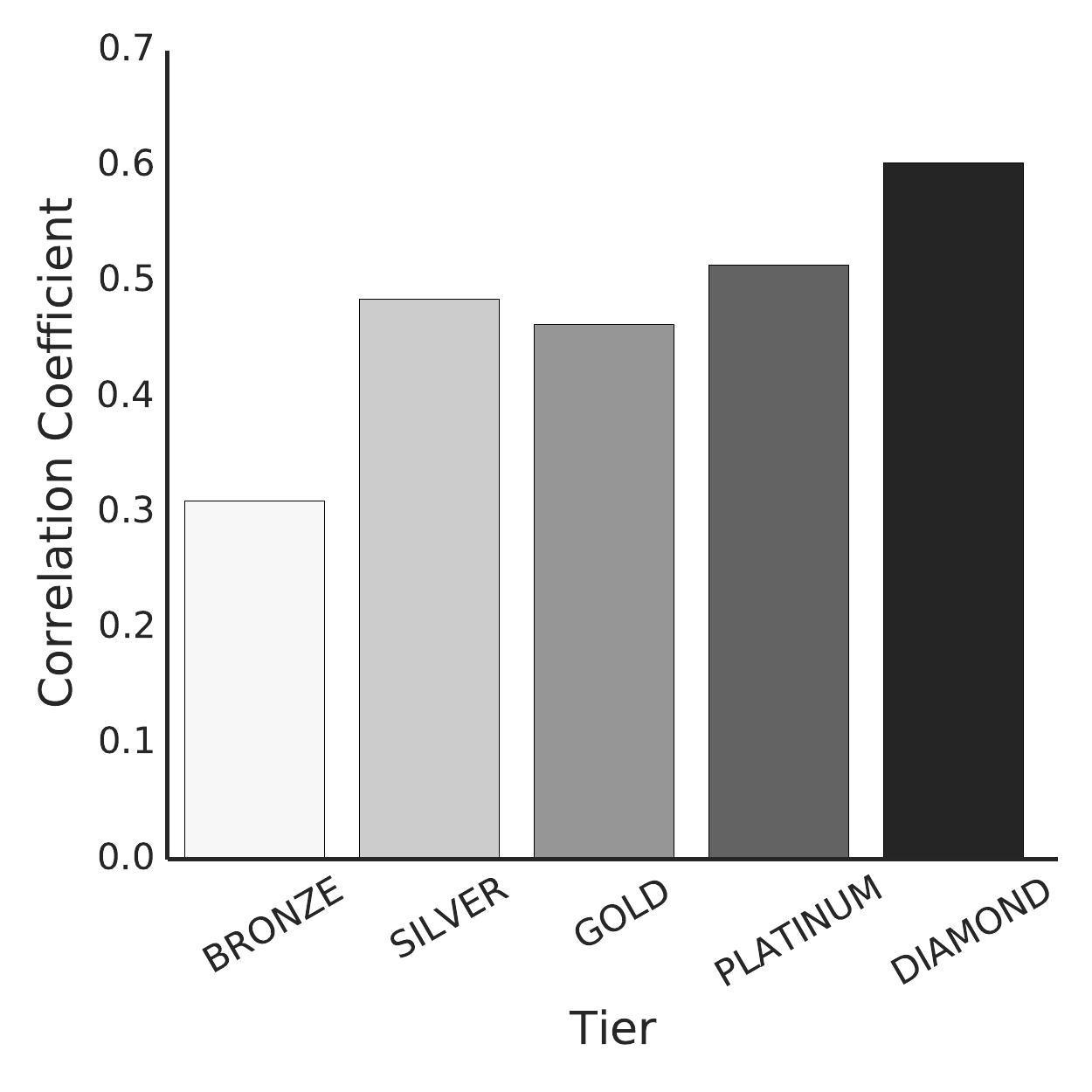}
   \caption{Correlations between the feature distances and subjective distances in the champion similarity spaces by survey respondents' tiers. Expert players in the diamond and platinum tiers have more accurate evaluations of champions' feature distances than non-expert players.}~\label{fig:poll_feature_corr}
 \end{figure}

There is an overall correlation of 0.503 between the feature and subject distances. However, subjective distances vary substantially between players in the different tiers. Figure~\ref{fig:poll_feature_corr} shows the correlations between the player responses from different tiers with the feature distances. Bronze players' subjective ratings of champion similarities are more weakly correlated with their ``true'' feature distance than the subjective ratings of Diamond players. This variability reflects the intuition that higher-tier players understand the champion characteristics better than the lower-tier players. In other words, a major component of player skill may be knowledge about champion similarity and how to design teams to maximize congruency. Because of this skill-based variability, we do not use the subjective distance in subsequent analyses. 

%While sampling higher level players for surveying will yield more reasonable champion distance compared to using all players, this result will be biased towards high-level players and will be unfair for low-level players when analyzing their team compositions. For these reasons, we used the survey distance only as a comparison to the feature distance.

%Based on the intercorrelations result and considering objectiveness, we select feature distance as the final candidiate for measuring distance between champions.

%\begin{table}[t]
%\centering
%\begin{tabular}{l | ccccc}
%& B & S & G & P & D \\ \hline
%\multicolumn{1}{c |}{Corr. Coef} & \multicolumn{1}{r}{0.31} & \multicolumn{1}{r}{0.48} & %\multicolumn{1}{r}{0.46} & \multicolumn{1}{r}{0.51} & \multicolumn{1}{r}{0.60}
%\end{tabular}
%\caption{Correlations between the feature distances and subjective distances in the champion similarity spaces by survey respondents' tiers. Expert players in the diamond and platinum tiers have more accurate evaluations of champions' feature distances than non-expert players.}~\label{tbl:poll_feature_corr}
%\end{table}

%\begin{table}[t]
%\centering
%\resizebox{0.45 \textwidth}{!}{%
%\begin{tabular}{  l | r  r  r }
%& \multicolumn{1}{c}{Feature} & \multicolumn{1}{c}{Co-occurrence} & \multicolumn{1}{c}{Subjective} \\ \hline
%Feature & 1 & 0.284 & 0.503  \\ 
%Co-occurrence & 0.284 & 1 & 0.192 \\ 
%Subjective & 0.503 & 0.192 & 1 \\  
%\end{tabular}}
%\caption{The symmetric correlations of champion distances calculated by champion features, champion co-occurences, and subjective similarity.}~\label{tab:intercorrelations}
%\end{table}

\subsection{Individual-level Features}
To formalize the proficiency-congruency dilemma, we define two constructs at the individual player level. \textit{Proficiency} is the similarity between a player's current and most-played champion and \textit{generality} is the variability or concentration of a player's champion choices. 

\subsubsection{Proficiency}

There are currently 126 champions in \textit{LoL} that a player can select for each match, each of which has its own abilities and roles. Because it is both expensive and unlikely for any user to develop expertise playing every champion, users focus on playing a subset of champions that reflect their particular play style or enjoyment. However, players cannot always select their most preferred champion because other players in the team may want to play a champion with similar abilities. 

We define \textit{proficiency} for player $u$ in match $m$ for the selected champion $c_{u,m}$ as follows:%\vspace{-.25em}
\begin{equation}\label{eq:proficiency}
Proficiency(u, m) = cosine\_sim(\hat{c}_u, c_{u,m}),
\end{equation}
%\vspace{-.5em}
where $\hat{c_u}$ is the most-picked champion of player $u$ during the 2015 season. Hence, if the player selects her most-picked champion, her proficiency for the match would be 1, and if she selects a champion with features that are in every way the opposite of the champion she typically plays, her proficiency for the match would be the minimum value -1.

\subsubsection{Generality}
Player $u$'s \textit{generality} is represented by the entropy of the player's seasonal champion selection distribution, which is defined as follows:%\vspace{-.25em}
\begin{equation}\label{eq:generality}
Generality(u) = - \sum_{i}{\frac{C_{u,i}}{C_{u}}\log{\frac{C_{u,i}}{C_{u}}}},
\end{equation}
%\vspace{-.5em}
where $C_{u, i}$ is the count of matches in which user $u$ selected champion $i$, $C_{u}$ is the total number of matches that $u$ played during the 2015 season, and the summation is over all champions. The generality metric is match-independent, and players who play multiple champions evenly would have higher generality values than players who focus on a few champions.

\subsection{Team-level Features}
We define two team-level constructs and describe other team-level variables that could impact performance. A team's \textit{congruency} is the count of feature-space clusters (from Figure~\ref{fig:champ_distance}) represented on the team and a team's \textit{diversity} is the average pairwise distance of champion similarities. 

\subsubsection{Team Congruency}
Team congruency captures how each team fulfills multiple functional requirements necessary for team victory. Team congruency is based on the functional clusters we found in Figure \ref{fig:champ_distance} that align with the five positions in a LoL team. We count how many of those functional clusters are fulfilled by the selected champions on a team. 
%Thus, when multiple selected champions are in same cluster, the feature value for that team will be low. 
The mathematical formulation of congruency for team $t$ is as follows:%\vspace{-.25em}
\begin{equation}\label{eq:congruency}
Congruency(t) = \sum_{i=1}^5 f_i,
\end{equation}
where $i$ denotes each functional cluster, and $f_i$ is a binary feature with the value 1 when functional cluster $i$ is fulfilled (i.e., one of the champions in cluster $i$ is in team $t$) and 0 otherwise. %Similar to the K-means clusters, each of the champion's PCA components, constructed for measuring feature distance, captures how the champion is designed to fulfill certain roles. The sum of the minimum and maximum PCA components for the 5 champions Thus, we look at the sum of the minimum and maximum values of the 5 champion PCA components.\brian{Can you say more --- is this the PCA from the feature distance space?} \jooyeon{I tried to add some more details. Let me know if you need more.}

\subsubsection{Team Diversity}
We define the team diversity of team $t$ for match $m$ by averaging the pairwise distances of team members' champions in the feature-based champion similarity space:%\vspace{-.25em}
\begin{equation}\label{eq:diversity}
Diversity(t, m) = {\begin{pmatrix} S \\ 2  \end{pmatrix}}^{-1} \sum_{uv(u \neq v)}{cosine\_dist(c_u, c_v)},
\end{equation}
%\vspace{-.5em}
where $S$ is the team size (always 5 in this case) and $c_u$ and $c_v$ are the vectorized expressions of the champions that users $u$ and $v$ selected for match $m$ (we simplified the expression by abbreviating $m$). Following previous research on team diversity and performance~\cite{stirling_general_2007}, we calculate the pairwise distance to represent group diversity. When champions selected by team members are similar, the team diversity value will decrease. We also use the team's minimum and maximum distances for win-lose prediction. 

\subsection{Team Assignment Features}
Team assignment category includes features that impact performance but are uncontrollable by team members. 

\subsubsection{Starting Location}
Starting location is a binary variable that tells whether a team starts at the top or bottom; since The Summoner's Rift is not perfectly symmetric, teams that start on the bottom win 50.8\% of the time.

\subsubsection{Background Diversity}
To capture the degree of overlap among team members' proficient champions, we consider the diversity of players' most-picked champions as background diversity (BD). This feature calculates the pairwise champion distance between team members' most-picked champions ($\hat{c_u}$ from Equation~\ref{eq:proficiency}), not their selected champion in a match:%\vspace{-.25em}
\begin{equation}\label{eq:background_diversity}
BD(t, m) = {\begin{pmatrix} S \\ 2  \end{pmatrix}}^{-1} \sum_{uv(u \neq v)}{cosine\_dist(\hat{c_u}, \hat{c_v})},
\end{equation}
%When this team value is higher than that of the opponent's, the team will be advantageous because they do not need to compromise their champion selection more than their opponent members. We also included the maximum and minimum pairwise distances of the proficient champions into our feature set.
% \brian{I don't understand the difference --- is this for the most-picked champions across all matches and matches?} \jooyeon{I modified a little bit. 5 team members have their most-pick champions respectively (these champions are match-independent), so you calculate the pairwise distance of those 5 post-pick champions.}

\subsection{Win-loss Classification}
Using only the team design features described above, we classify whether each team will win or lose the match. We use logistic regression classifiers trained for each tier and region with five-fold cross-validation to predict each game's binary win/loss label. Using the full model of team design features, we achieve an average classification accuracy of 58.4\%. In traditional machine learning approaches this level of performance would be considered low, but two factors are important to keep in mind. First, \textit{LoL} uses a matchmaking system to assign players to create balanced teams with an expected win rate of 50\%~\cite{riot_matchmaking}. Second, our classification model uses only team design features. Including in-game performance metrics (kills, deaths, gold, \textit{etc.}) would greatly improve the performance of the classifier, but our goal is to show that players' team design choices before a match starts has substantial influence on team performance.

We can roughly benchmark the influence of team design on win rates by comparing it to changes in win rates for other strategies analyzed and discussed in the player community. First, even small departures of $\sim$3\%, such as the starting location bias, are widely discussed in the player community~\cite{lejazq_blue_2014}. The win rate for the first team to kill an opponent's champion (``First Blood'') rises to $\sim$60\% and is acknowledged as a critical factor for gaining cumulative advantage (``snowballing'') in a match~\cite{HawatLoL_firstblood_2014}. Thus, our results suggest teams that jointly maximize proficiency and congruency start the game with a substantial advantage comparable to more popular strategies such as winning early objectives~\cite{snowballing_2013,graphs_red_blue}.

In the third column of Table \ref{tbl:features}, we show the win-loss classification accuracy using each feature alone. The result shows that \textit{ individual proficiency} is the most powerful feature for classification, followed by the starting position binary variable, team congruency feature, and team diversity feature. Interestingly, player's degree of champion generality and background diversity show relatively weak classification performance compared to other features. Thus, features that are derived match-wise by players' decision and interaction as a virtual team are more important for win-loss result than the features that are not match-specific, for example, player's champion selection history or starting position.

%For win-loss classification, we compute the proficiency of the first through fifth picks, and also the average value of those 5. The Distribution column of Table~\ref{tbl:features} plots the histograms of each feature value binned by ranked tier, from Bronze V (the lowest, on the left) to Challenger I (the highest, on the right). 

\subsection{User Interviews}
Twenty LoL players in Korea were recruited for user interviews, ten elite players (at least Platinum tier) and ten novice players (at most Silver tier) via postings to online communities to play the game in lab in exchange for approximately 20 USD. Participants played three games as co-located teams of five under different communication conditions (none, text, voice) and we conducted a focus group interview for one hour and asked how they selected the champions for those games. The focus groups interviews were recorded and transcribed in Korean and coded using an inductive approach~\cite{elo2008qualitative, thomas2006general}. %Two of the investigators are native Korean speakers and carried out the interview and analysis.

%%%%%%%%%%%%%%%%%
% Results %
%%%%%%%%%%%%%%%%%

\begin{table}[t]
\centering
\label{my-label}
\resizebox{0.45 \textwidth}{!}{%

\begin{tabular}{llr}
 & {\bf Distribution} & {\bf Classification} \\
  & {\bf Across Tiers} & {\bf Accuracy} \\
  \hline
   {\bf Using all features} &  & {\bf 58.4} \\ \\

{\bf Individual-level features} &   & {\bf 57.0}  \\ \hline
% {\bf Proficiency features} &    &  {\bf 57.0} \\
Mean proficiency of players     &          \raisebox{-5px}{\includegraphics[width=0.1\textwidth]{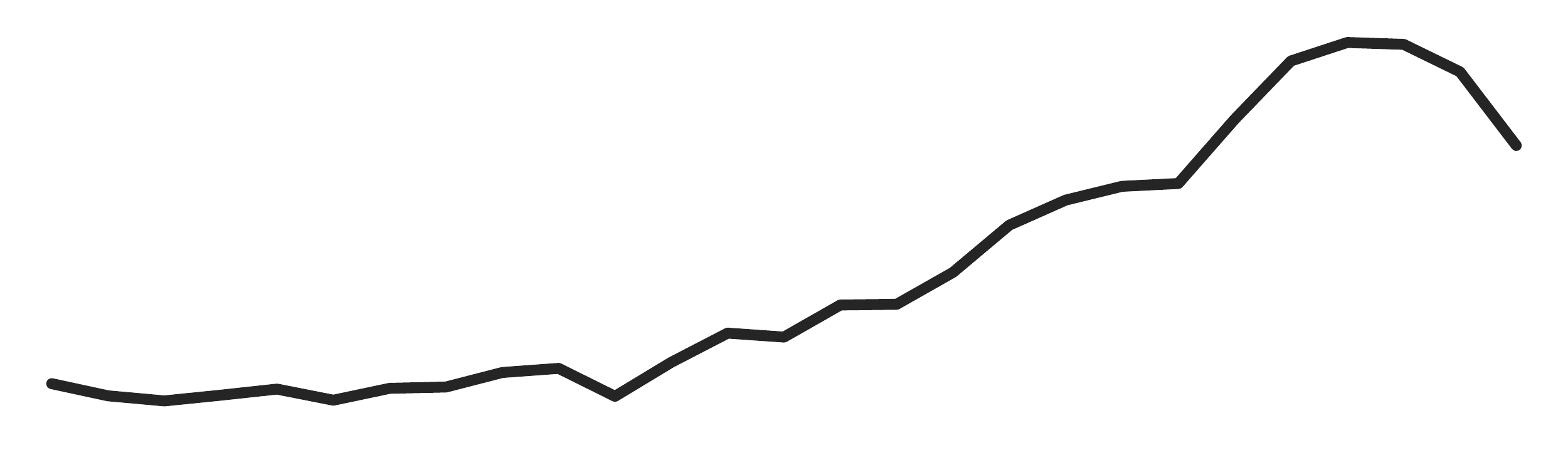}}      &      56.9    \\  
% 1st pick's proficiency &    \multirow{5}{*}{\raisebox{-5px}{\includegraphics[width=0.1\textwidth]{figures/test3}}  }  &     53.9  \\
% 2nd pick's proficiency          &                    &  53.8 \\
% 3rd pick's proficiency          &                    &  53.9  \\
% 4th pick's proficiency          &                    &   53.8   \\
% 5th pick's proficiency          &                    & 53.6    \\ \hline

% {\bf Generality features} &    & {\bf 51.7} \\
Mean champ. generality      &  \raisebox{-5px}{\includegraphics[width=0.1\textwidth]{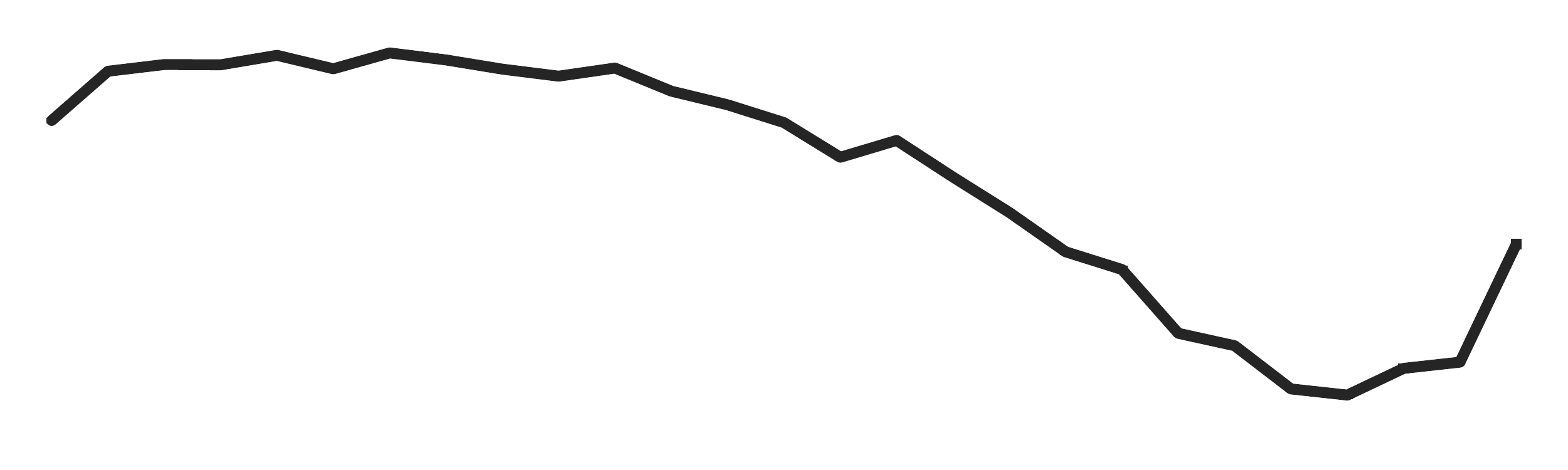}}      & 51.2\\ \\
% 1st pick's champ. generality   &\multirow{5}{*}{\raisebox{-5px}{\includegraphics[width=0.1\textwidth]{figures/test5}}  } & 51.3\\
% 2nd pick's champ. generality&                    & 51.0 \\
% 3rd pick's champ. generality&                    &  50.8 \\
% 4th pick's champ. generality&                    & 50.6 \\
% 5th pick's champ. generality&                    & 50.6 \\ \\
{\bf Team composition features} &                    &    {\bf 52.6}       \\ \hline
Team congruency            & \raisebox{-5px}{\includegraphics[width=0.1\textwidth]{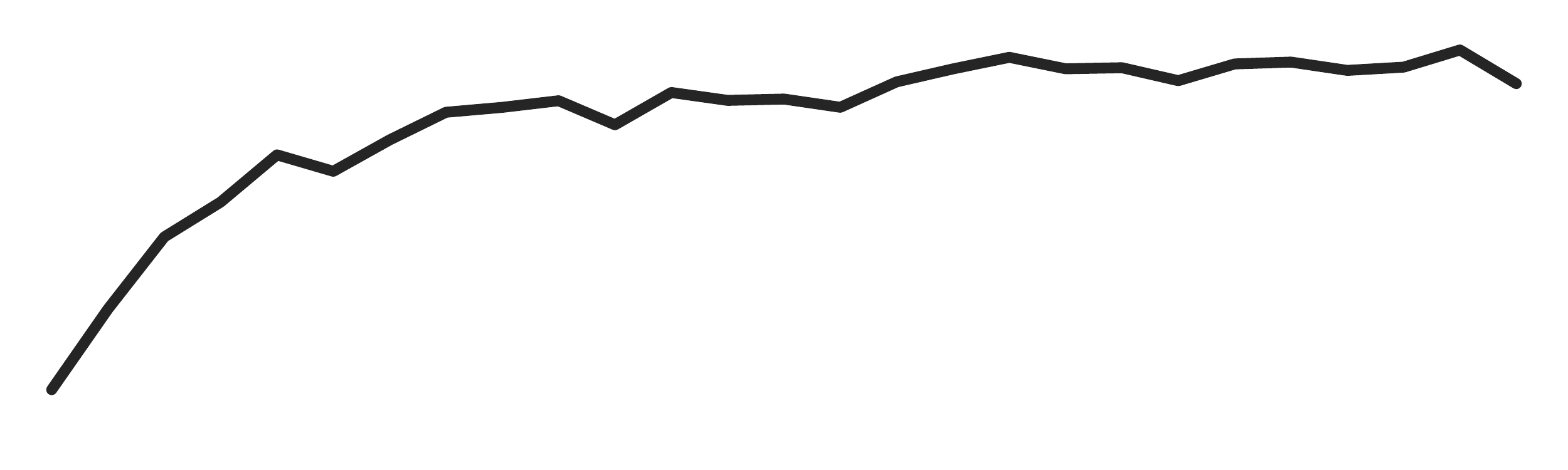}}      & 51.9\\
%Sum of min. PCA comp.      &     \raisebox{-5px}{\includegraphics[width=0.1\textwidth]{figures/test9}}  &51.5\\
%Sum of max. PCA comp.       &   \raisebox{-5px}{\includegraphics[width=0.1\textwidth]{figures/test10}}& 50.7  \\
Team diversity                 &\raisebox{-5px}{\includegraphics[width=0.1\textwidth]{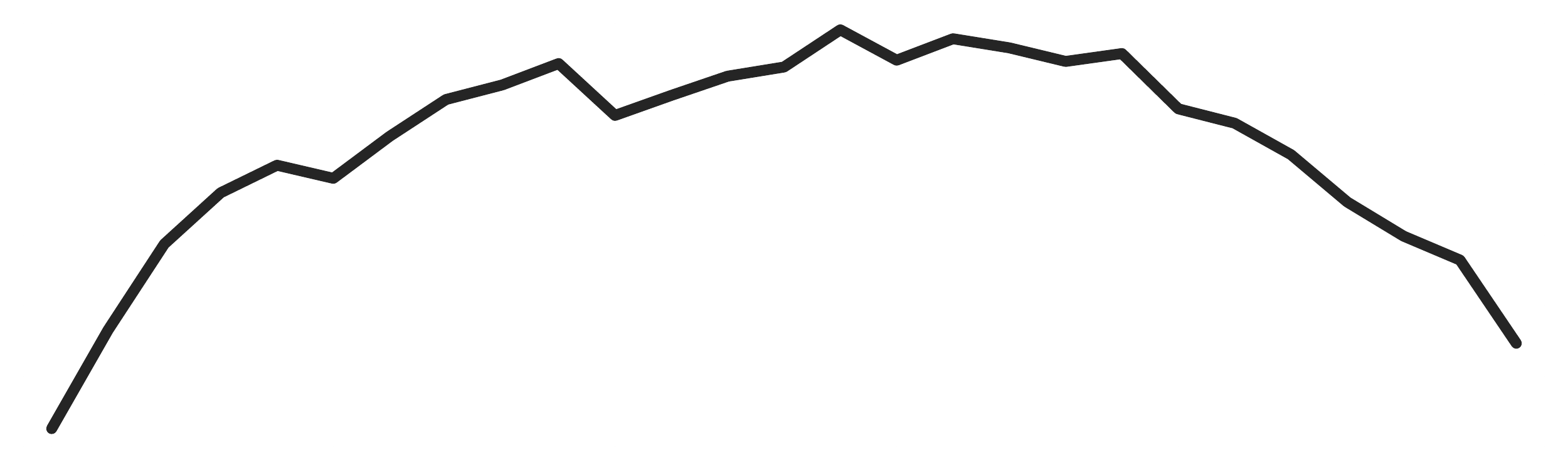}}     &51.6\\
Min. champ. distance            &            \raisebox{-5px}{\includegraphics[width=0.1\textwidth]{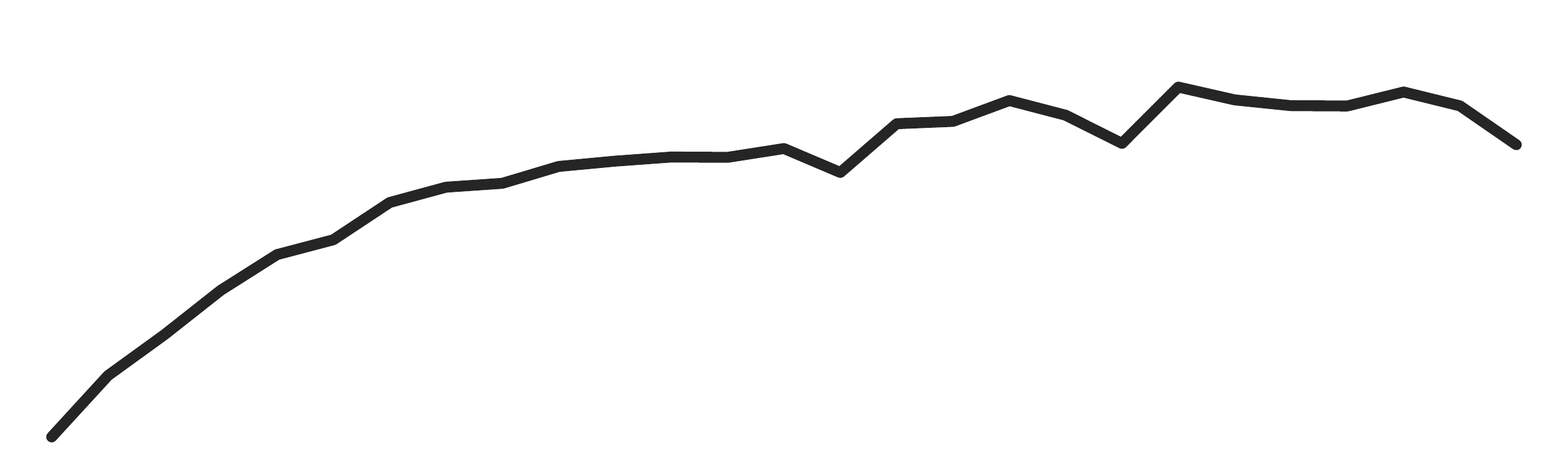}} & 51.0  \\
Max. champ. distance            &           \raisebox{-5px}{\includegraphics[width=0.1\textwidth]{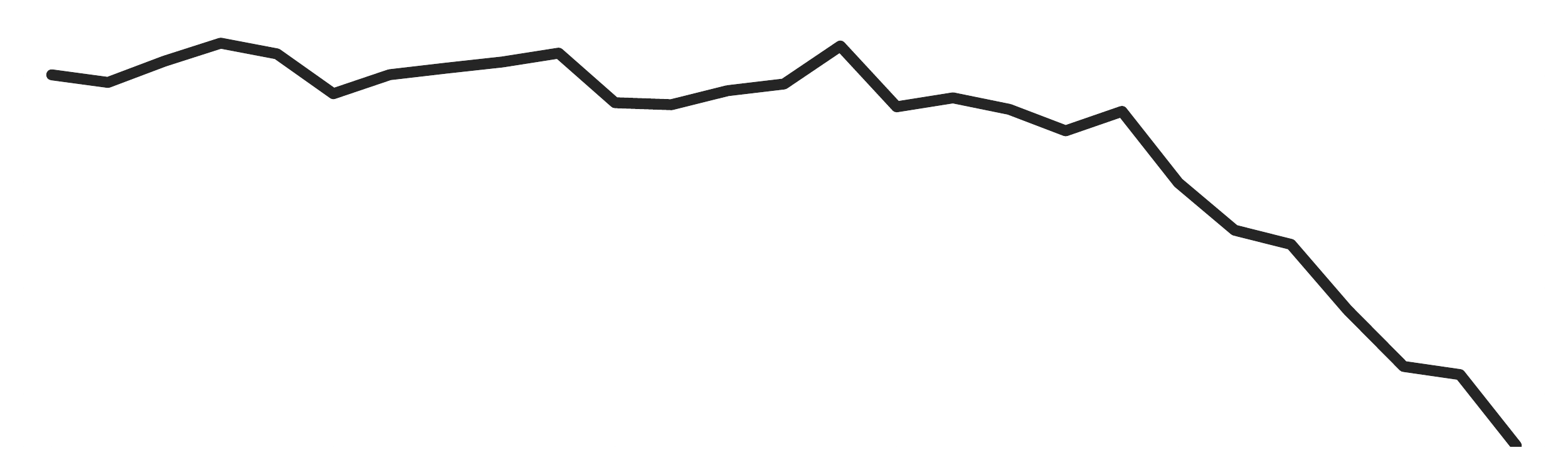}} & 50.7 \\ \\

{\bf Team assignment features}  &                     &  {\bf 53.6} \\ \hline
Starting Position               &               \raisebox{-5px}{\includegraphics[width=0.1\textwidth]{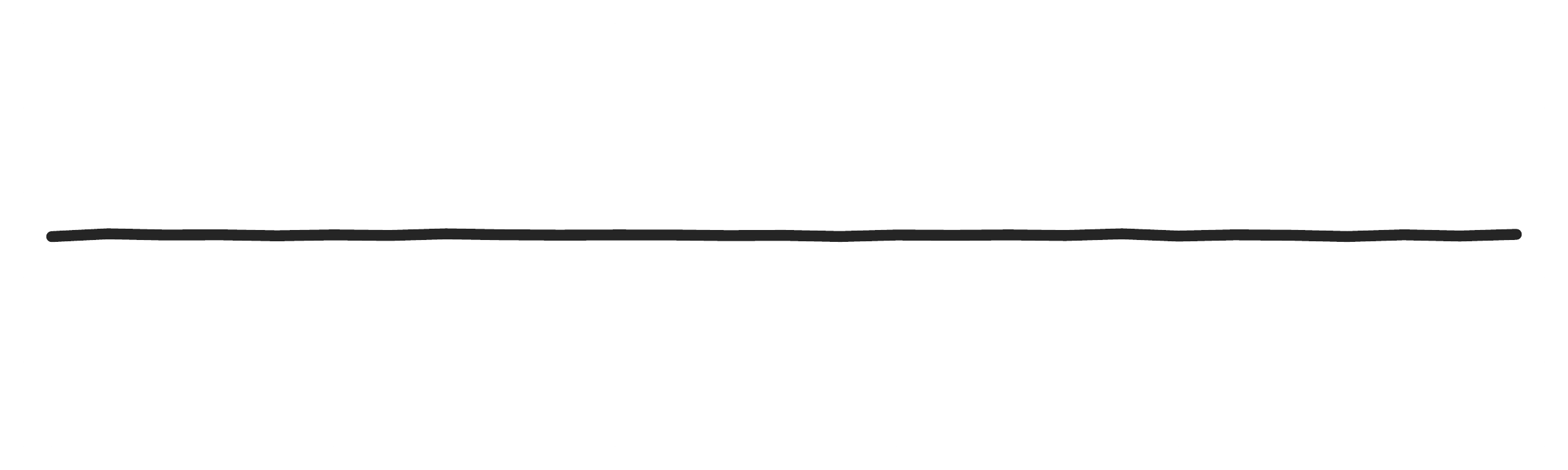}}       & 53.5\\
Background diversity &    \raisebox{-5px}{\includegraphics[width=0.1\textwidth]{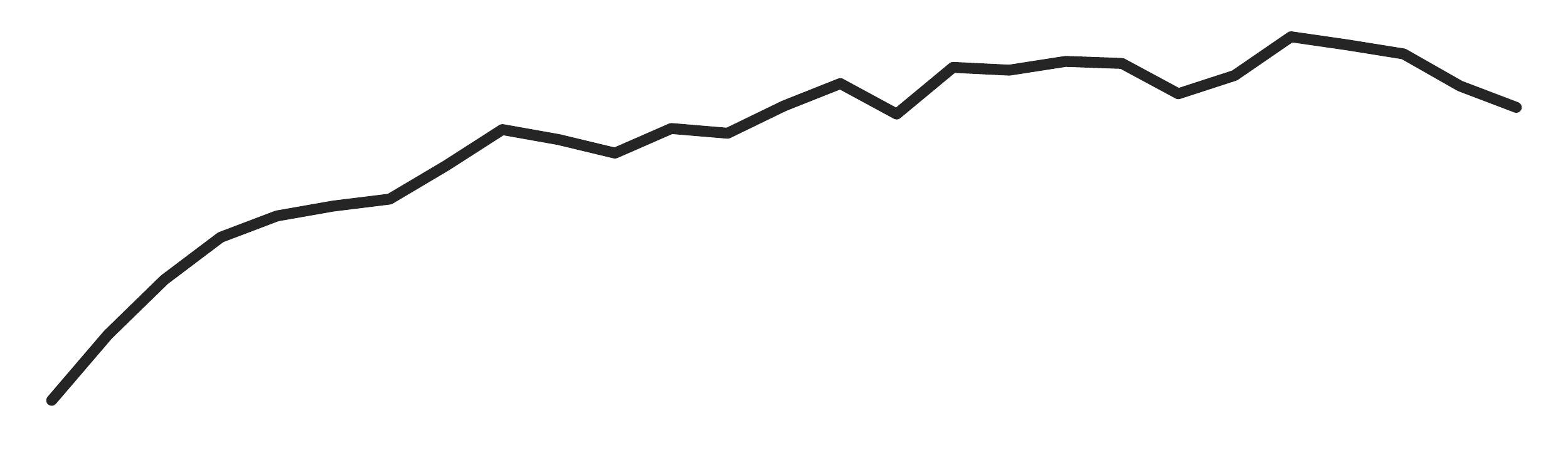}}  & 50.8\\
Min background diversity   &       \raisebox{-5px}{\includegraphics[width=0.1\textwidth]{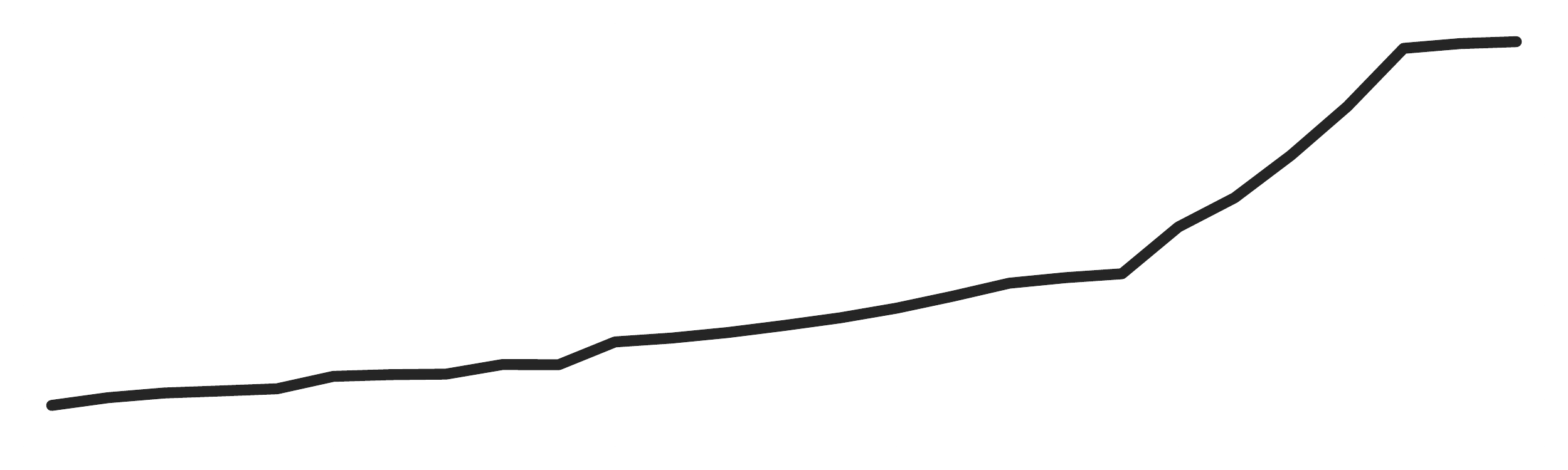}}  &50.9 \\
Max background diversity   &     \raisebox{-5px}{\includegraphics[width=0.1\textwidth]{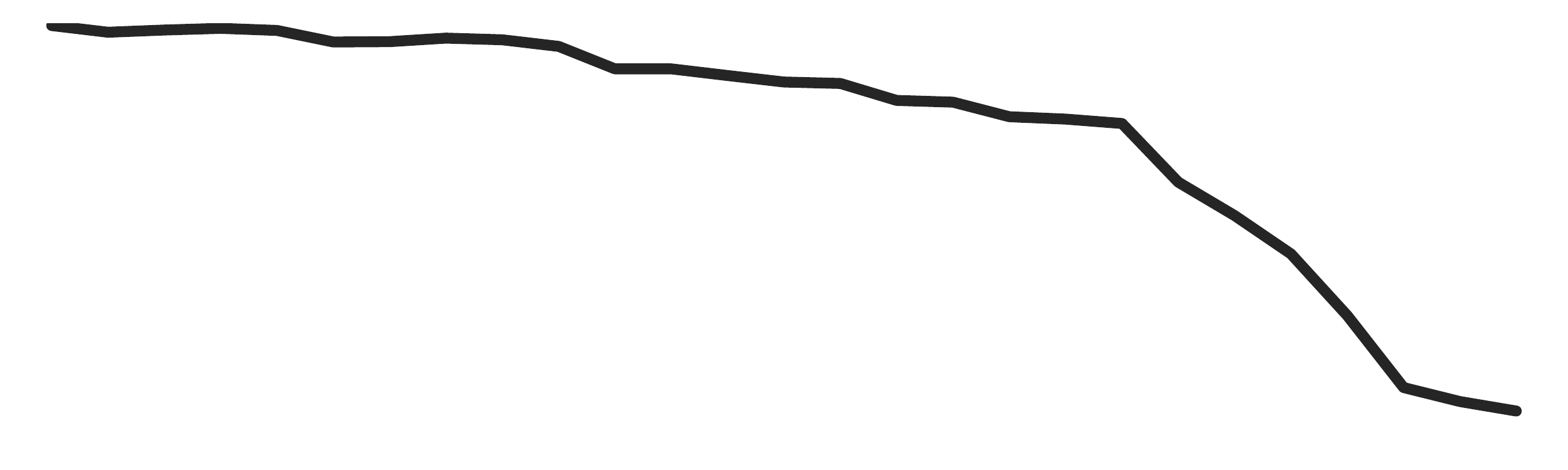}}  & 50.4 
\end{tabular}
}
\caption{Individual and team features used for win-loss classification, their distributions across the tiers (Bronze I on left and Champion on right), and the average accuracy from the binary classifier using the features or ensembles of features.}\label{tbl:features}
\end{table}

\begin{figure*}[tb]
 \centering
 \minipage{0.6\columnwidth}
    \centering
  	\includegraphics[width=\linewidth]{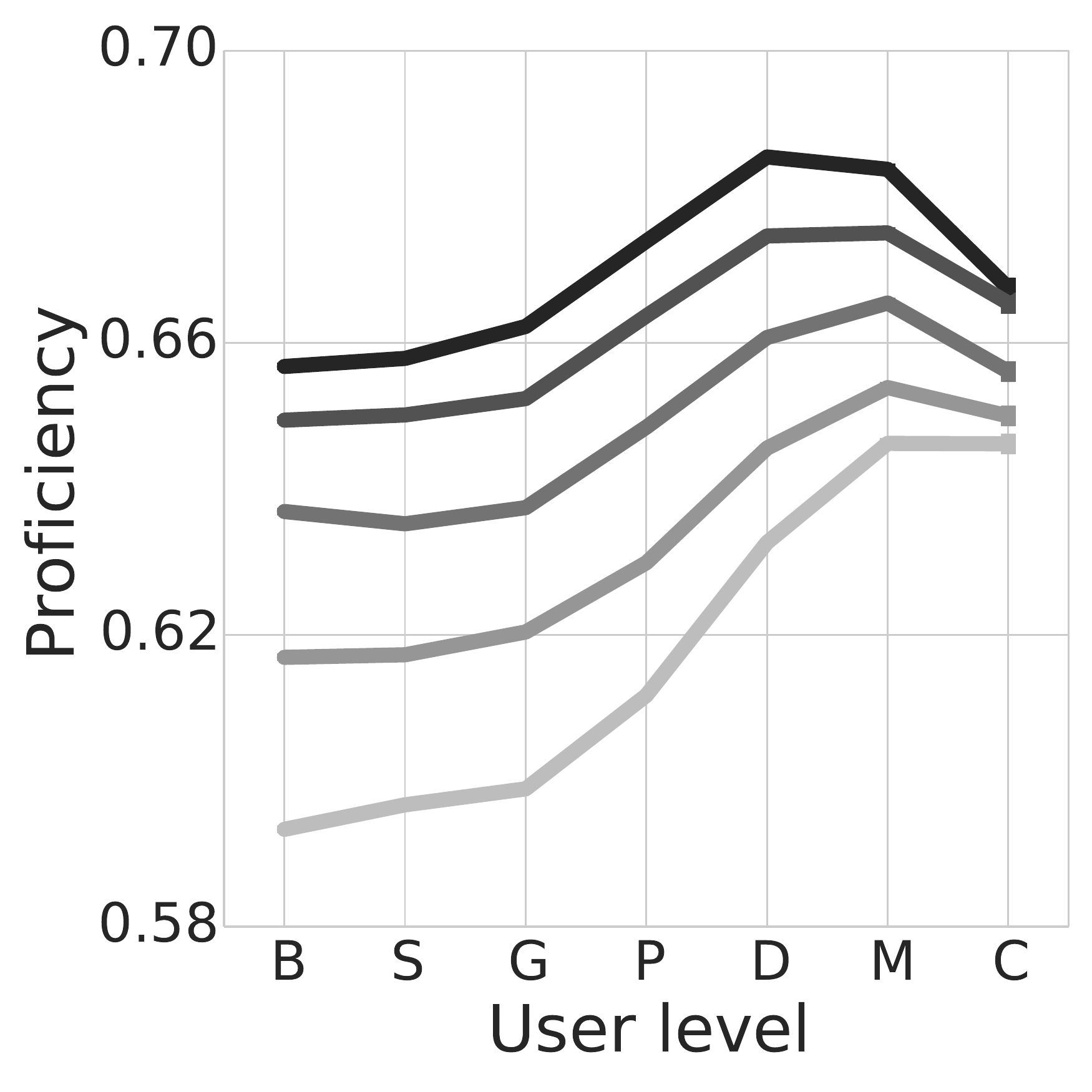}\\(A)
 \endminipage\hfill
 \minipage{0.6\columnwidth}
    \centering
  	\includegraphics[width=\linewidth]{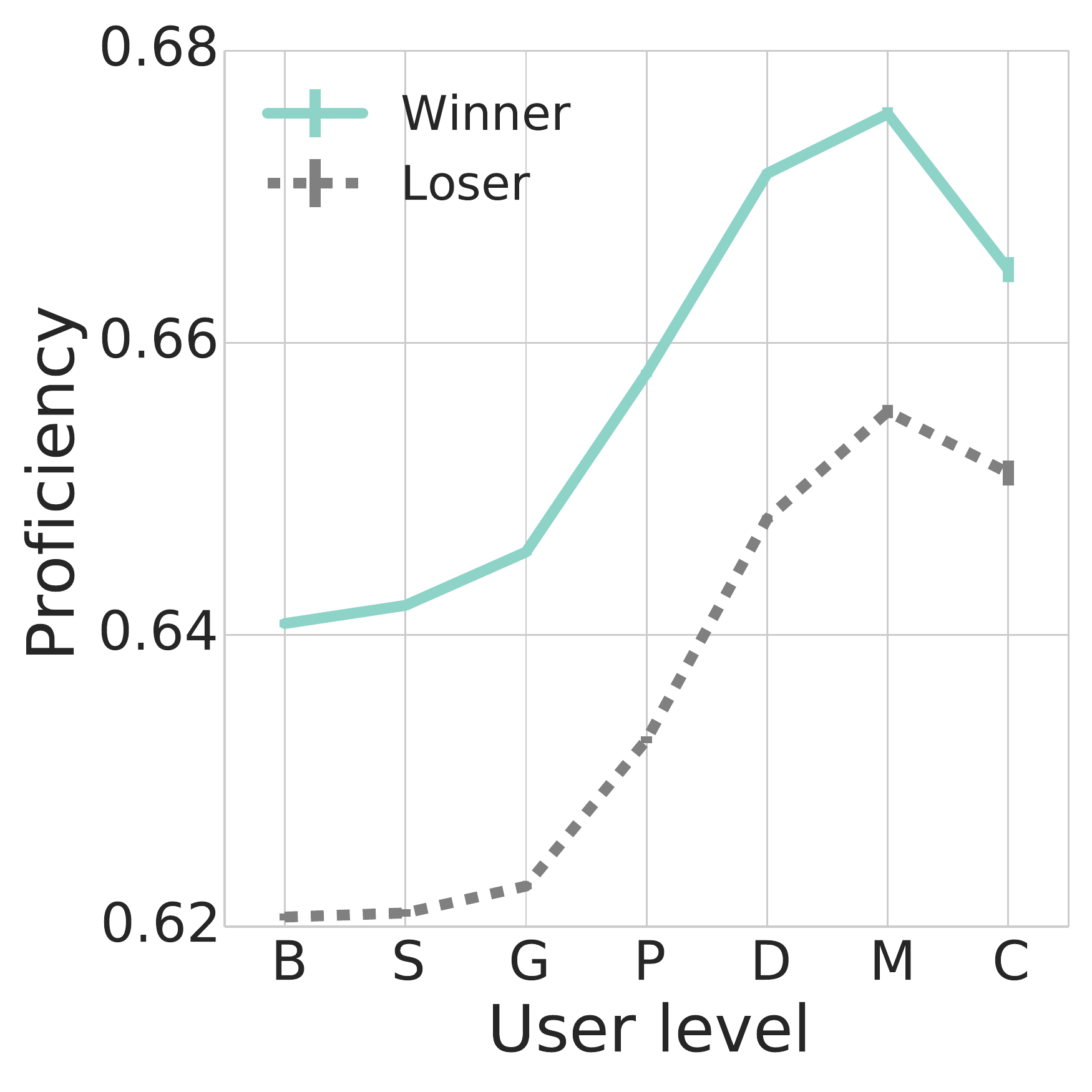}\\(B)
 \endminipage\hfill
 \minipage{0.6\columnwidth}
    \centering
    \includegraphics[width=\linewidth]{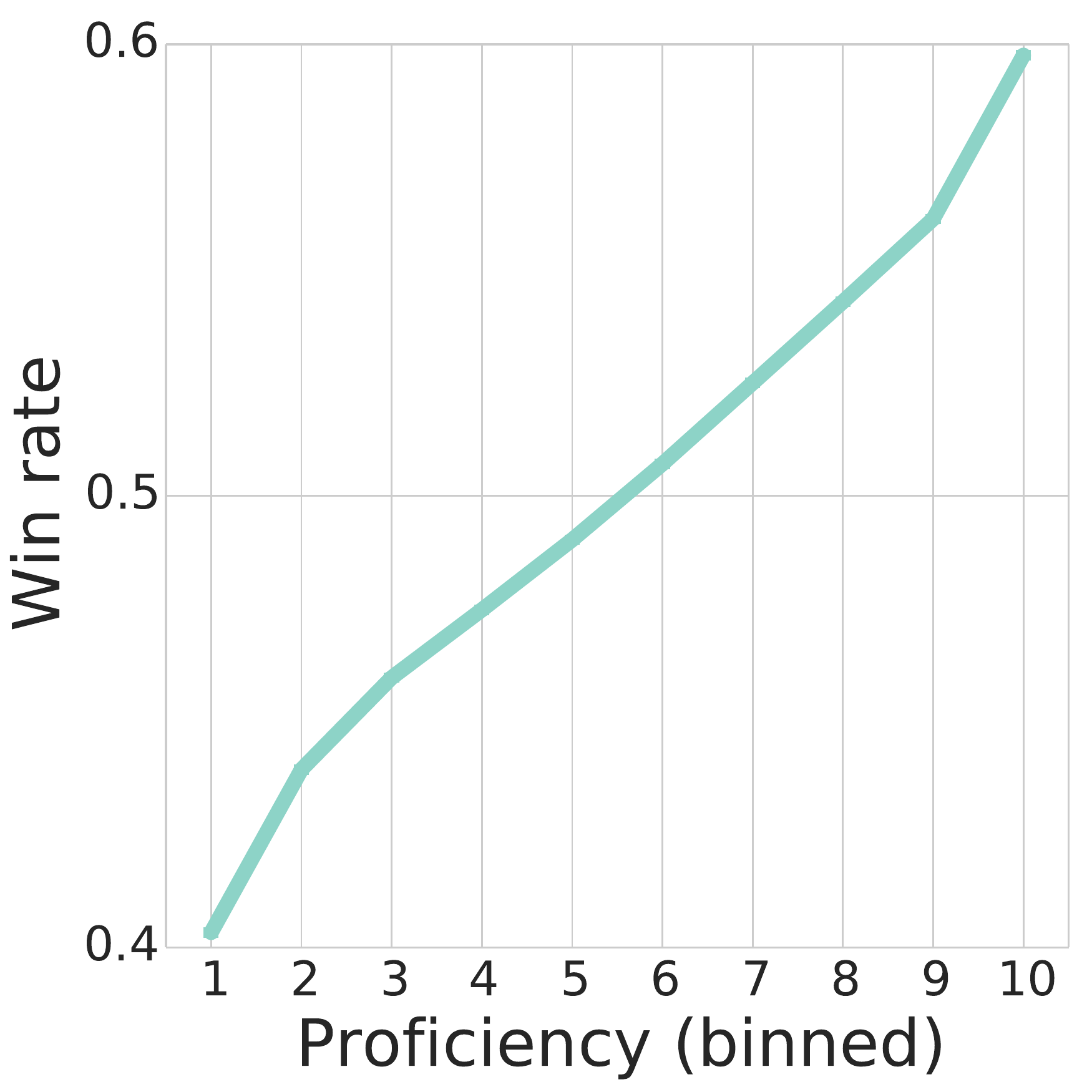}\\(C)
 \endminipage\hfill
 \caption{(A) The variation in the first (darkest line) to fifth (lightest line) picks' proficiency by ranked tier. (B) The variation in average team proficiency by ranked tier. (C) The relationship between relative proficiency and team performance.}\label{fig:proficiency}
\end{figure*}

\section{H1: Proficiency increases performance}
Hypothesis 1 predicted that teams with members using champions they play frequently will perform better than teams with members using champions they play infrequently. The average accuracy scores show that the player-level proficiency scores increased win prediction accuracy to 57\%, which makes this the most important set of features in the full model. Individual-level generality features produced only a minor increase in accuracy to 51.2\%. Hypothesis 1 is supported because individual-level features of proficiency and generality both predict team performance. %We explore these results in more quantitative and qualitative detail below.

\subsection{Quantitative findings}
The feature distributions in Table~\ref{tbl:features} for Individual-level features show a non-monotonic relationship between player expertise (ranked tiers on the $x$-axis) and players' average proficiency and generality values ($y$-axis). Players' proficiency increases and generality decreases as the tier goes from Bronze to Diamond, but proficiency decreases and generality increases significantly from Master to Challenger. The first trend suggests players are able to rise through the rankings by mastering a single character through repeated play. But the reversal of proficiency and generality among players in the Master and Challenger tiers suggests they employ a different strategy, either by voluntarily selecting a less-proficient champion to help the team or having to react to their ``first choice'' being removed from play by the other team. 

There is also an interaction effect with the proficiency gap between fifth picks for novice players' versus elite players (Figure \ref{fig:proficiency}A): the difference in proficiencies between the first and fifth picks among novices is very large while the difference in proficiencies among elite players is much smaller. This suggests elite players use pick-order to trade champions and team members' proficieny more than novices.

In Figure \ref{fig:proficiency}B, we observe that the mean proficiency values for winning teams (green) are significantly higher than the mean proficiency values for losing teams (grey) across tiers. The within-tier difference between mean proficiency values is likewise stable across tiers, suggesting that mean proficiency is a robust indicator of a team's likelihood of winning regardless of expertise. Figure \ref{fig:proficiency}C illustrates the relationship between individual proficiency and win rate by averaging team's proficiency scores relative to their opponents. Teams with greater relative proficiency see significant increases in win rates, exceeding 60\% in some cases.

%both mean proficiency and team diversity values are always higher for the winning teams than losing teams, regardless of tiers and divisions; although the values fluctuate, the gap between winning and losing teams are consistent for all user levels. Other powerful features for prediction also show the salient difference between the winning and losing teams. Thus, we confirm that our features for win-lose prediction is applicable to all user levels.
%In Table \ref{tbl:features}, we confirm that higher-tier players are more specialists than lower-tier players.

\subsection{Qualitative findings}
Our interviews with players corroborate our findings from above that elite players pick stable and safe champions to play in specific roles, especially on teams with members with which they are not familiar. 

\begin{quote}\textit{
``Usually I don't pick those champs, but I picked Morgana because it is stable... it grows steadily, and I can play steadily and will be brought to an end with steady plays. I wanted to play all-or-nothing style champs, but I made safer picks rather than that." (P9, Platinum 5)
}\end{quote}

\begin{quote}\textit{
``I pick what I can do well, except a counter-pick appeared in the opponent team... I only avoid those worst combinations. Picking the proficient champ is good for teamwork as well." (P1, Diamond 5) % (When I had to play a AD-carry) One of the worst bot duo was, Vayne and Soraka... 
}\end{quote}

\begin{quote}\textit{
``In my case, If I'm really good at a specific champ, and confident about understanding what I should do in various situations, I will pick that." (P7, Diamond 5)
}\end{quote}

Elite players think that \textit{``taking a role with low proficiency means losing the game''} and separate competence from enjoyment when choosing roles to play:

\begin{quote}\textit{
``You will realize which lane you are good at, independent of what you like, as your tier gets higher. If you play just what you like, you will probably lose because opponent player on your lane may very good at that lane... then you will lose." (P1, Diamond 5)
}\end{quote}

Novice players attributed team success more to luck or player proficiency than team-level processes.

%be inefficient to resolve the dilemma that leads to lower functional team diversity in limited time. Moreover, in some cases, some novice players think team diversity doesn't even matters:

\begin{quote}\textit{
``If you are Bronze tier, luck matters. you can raise your tier by picking a champ you're are good at, in Silver." (P12, Silver 4)
}\end{quote}
\begin{quote}\textit{
``On Silver, it's better to select the champ you are good at... such a perception is widespread among players." (P14, Silver 3)
}\end{quote}
\begin{quote}\textit{
``Winning the lane fight is important, it's better to pick a familiar champ." (P13, Silver 4)
}\end{quote}

Novice players' perceptions that expertise and luck were the most important factors for progresses up into higher ranked tiers is not matched to the same extent by elite players. They select champions that are ``safer'' or ``steadier'', reflecting a particular style of play and knowledge about how the champion progresses through the game. %These findings corroborate our quantitative results that player proficiency with a champion is a major determinant of team success. 

\section{H2: Congruency increases performance}
Hypothesis 2 predicted that teams composed of more congruent champions will perform better than teams with less congruency. The average cross-validated scores using only the team composition features increased the prediction accuracy to 52.6\%, a significant but modest lift relative to individual proficiency features. Hypothesis 2 is supported because team composition positively predicts team performance. %As before, we explore these results in more quantitative and qualitative detail below.

\subsection{Quantitative findings}
\begin{figure}[tb]
 \centering
 \minipage{0.5\columnwidth}
    \centering
  	\includegraphics[width=\linewidth]{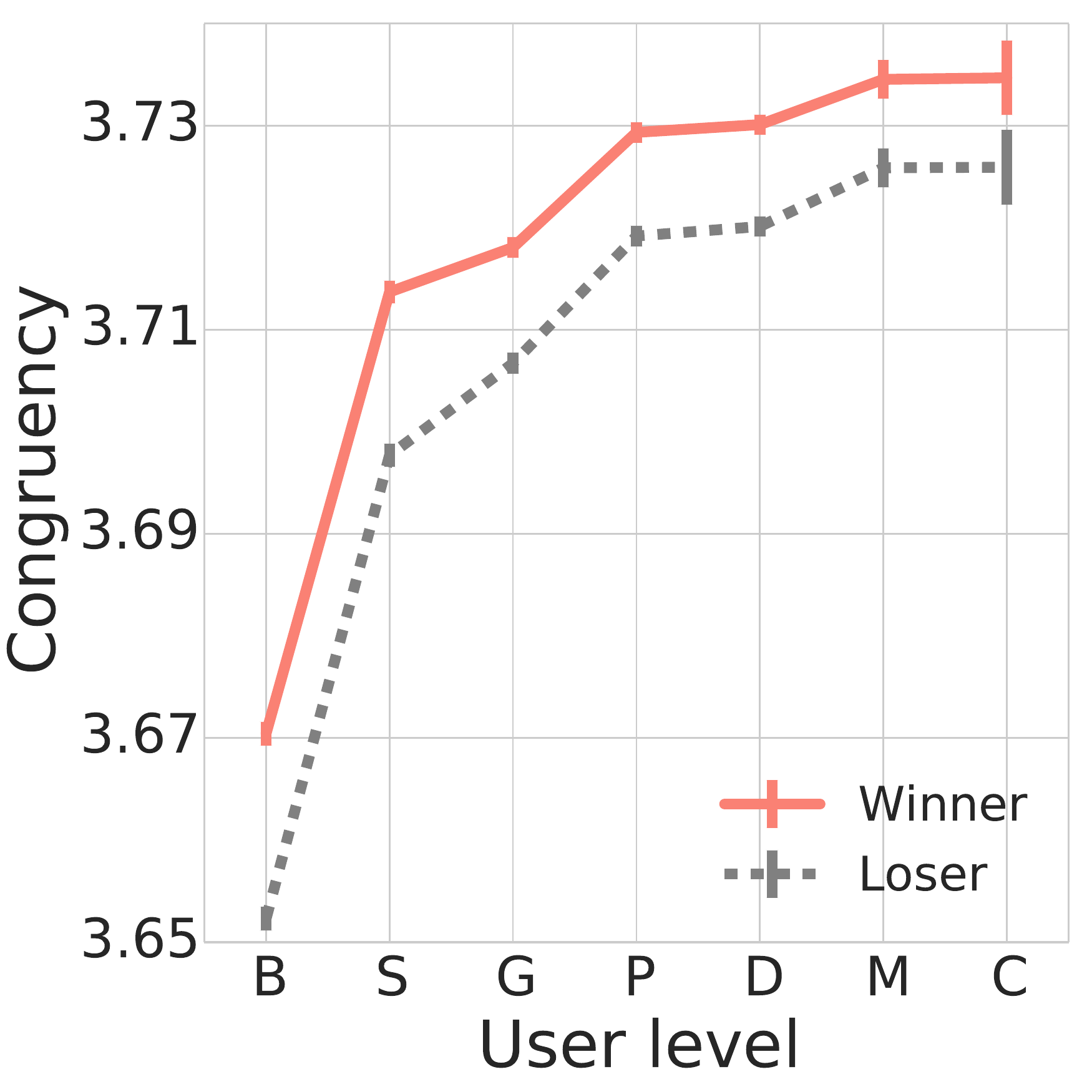}\\(A)
 \endminipage\hfill
 \minipage{0.5\columnwidth}
    \centering
    \includegraphics[width=\linewidth]{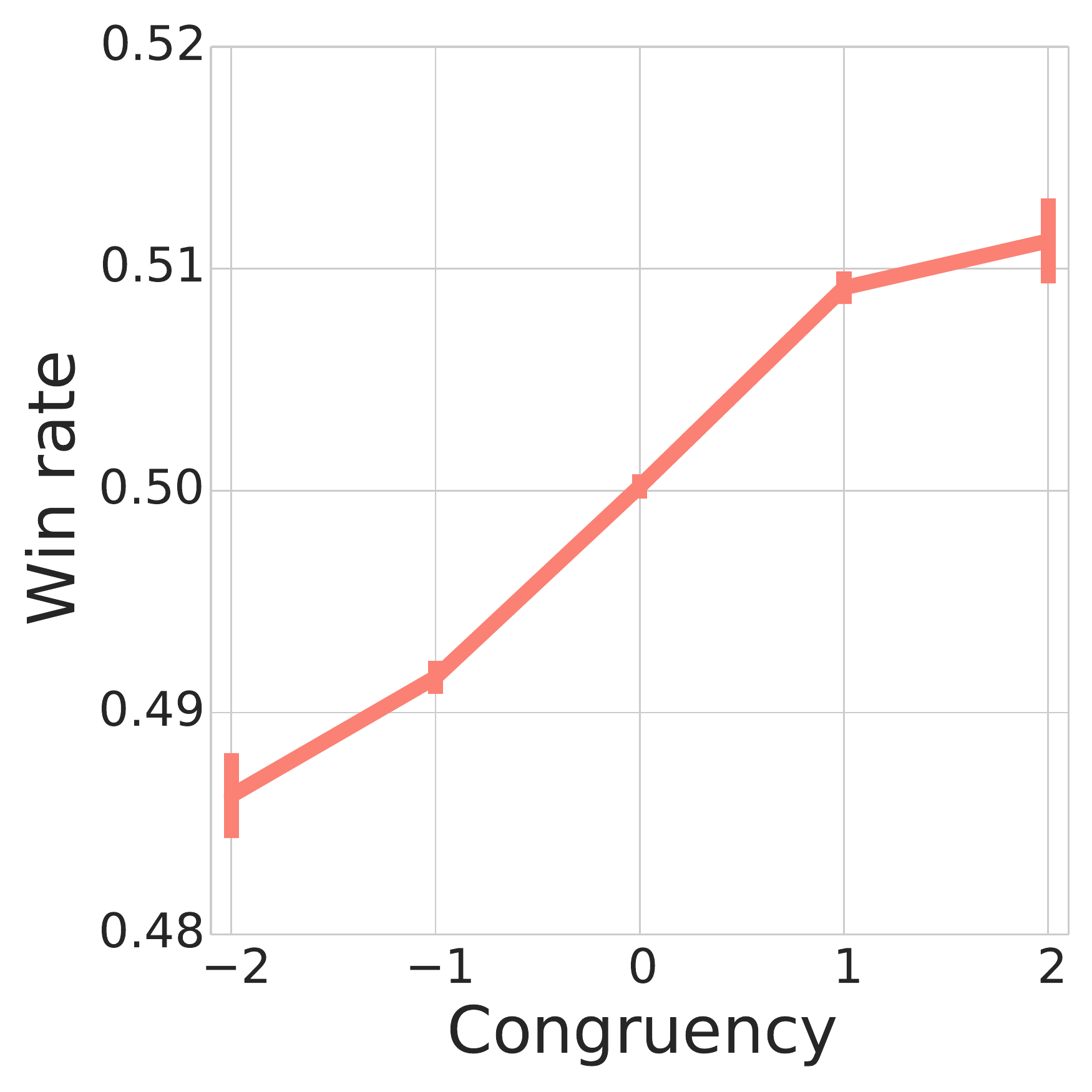}\\(B)
 \endminipage\hfill
 \caption{(A) The variation in average team congruency by ranked tier. (B) The relationship between relative congruency and team performance.}\label{fig:congruency}
\end{figure}

Table~\ref{tbl:features} shows the distributions for team congruency and diversity under ``Team composition features'' across tiers. Congruency captures whether all five clusters from Figure~\ref{fig:champ_distance} are represented and diversity is the average pairwise feature distance between champions. We see a sharp increase in congruency from Bronze to Gold players, which then stabilizes: novice players are different largely because they do not select champions to construct complementary teams. %The results also show that the sum of maximum PCA components is higher for elite teams than novice teams, while the sum of minimum PCA components is lower; for elite teams, their selected champions' principal component values are more dispersed than that of the novice teams.\brian{I'm still think reviewers/readers will struggle to understand the significance of this in terms of skill, teamwork, etc. How is max/min PCA different from diversity?}

The results for team diversity in Table~\ref{tbl:features} show an increase from Bronze tier to around Gold, but then decreases for elite players. This non-monotonicity is partially explained by the minimum and maximum champion distances. The minimum champion distance monotonically increases, signifying that higher level teams avoid functional overlap between champions, while the maximum champion distance monotonically decreases, signifying that the higher level teams avoid champion pairs that are extremely different from one another.

In Figure \ref{fig:congruency}A, we observe that both average team congruency values are always significantly higher for the winning teams than losing teams across all tiers and divisions. Again, the values of team congruency increase across tiers and the gap between winning and losing teams are consistent for all user levels. Figure \ref{fig:congruency}B illustrates the relationship between relative team congruency values and win rate. Teams' with higher congruency relative to their opponents have greater win rates. 

%Table \ref{tbl:features} shows that most-pick champions overlap more in lower-tier teams than higher-tier teams.

\subsection{Qualitative findings}
Elite players consider congruency an important factor and will pick champions to maximize congruency based on the trust of teammates' proficiency, even if they met for the first time in the game.

\begin{quote}\textit{
``As my tier gets higher, I trust my team members and pick what is needed... then even if I don't do something special, I can win." (P6, Platinum 2)
}\end{quote}

\begin{quote}\textit{
``When I was Silver, it is recommended to pick what I'm familiar with... But if you go up (to higher tier), combination is important, I'm always concerned about the champ combination (P10, Platinum 3)
}\end{quote}
%This is because elite players think that \textit{'taking a role with low proficiency means losing the game'}. P1, Diamond 5, reported:

%\begin{quote}\textit{
%"You will realize which lane (or a role) you are good at, independent to what you like, as your tier gets higher. If you play just what you like, you will probably lose because opponent player on your lane may very good at that lane (or a role)... then you will lose. (P1)"
%}\end{quote}

On the other hand, novice players may select roles more for fun than team congruency. 

\begin{quote}\textit{
``I have no main position... When playing solo-rank game, I have a few champions for each position I can do... Instead of playing 'Mid' again and again, just picking Thresh when I have to play a Supporter, picking Lucian when playing a AD-carry, Xerath for Mid... (P15, Silver 3)"
}\end{quote}

As the quantitative results above show, novice dispositions to ignore team composition factors and focus instead on champions or roles they enjoy playing have consequences for team performance. However, they nevertheless recognize the importance of team composition for success. %After a loss, P11 (Bronze 1) attributed the failure to the absence of any team member playing a ``tanking''.

\section{H3: Expert responsiveness to dilemma}
Hypothesis 3 predicted that elite players are more responsive during team design to the proficiency-congruency dilemma than novice players. Although the full model in Table~\ref{tbl:features} shows an increase in performance combining the individual and team-level features, we use our mixed methods results from below and conclude that Hypothesis 3 is supported.

\subsection{Quantitative findings}
%\begin{figure}[t]
%	\centering
%		  \includegraphics[width=0.9\columnwidth]{figures/div_slope_all.pdf}
%		  \caption{The correlations between team diversity and diversity of proficient champions (top) and individual proficiency (bottom). The second row encompasses mean individual proficiency (solid) as well as the 1st to the 5th pick's proficiency values (dashed, from dark to light). The correlations, both positive and negative, are stronger for lower-level players than for higher-level players. *, ** signifies that the regression slope is less than 0 with p-value $< 10^{-2}$ and $< 10^{-20}$ respectively (error bars are too small to be shown).}
		  \label{fig:div_coef}
%\end{figure}

\begin{figure*}[t]
\centering
 \minipage{.6\columnwidth}
    \centering
  	\includegraphics[width=\linewidth]{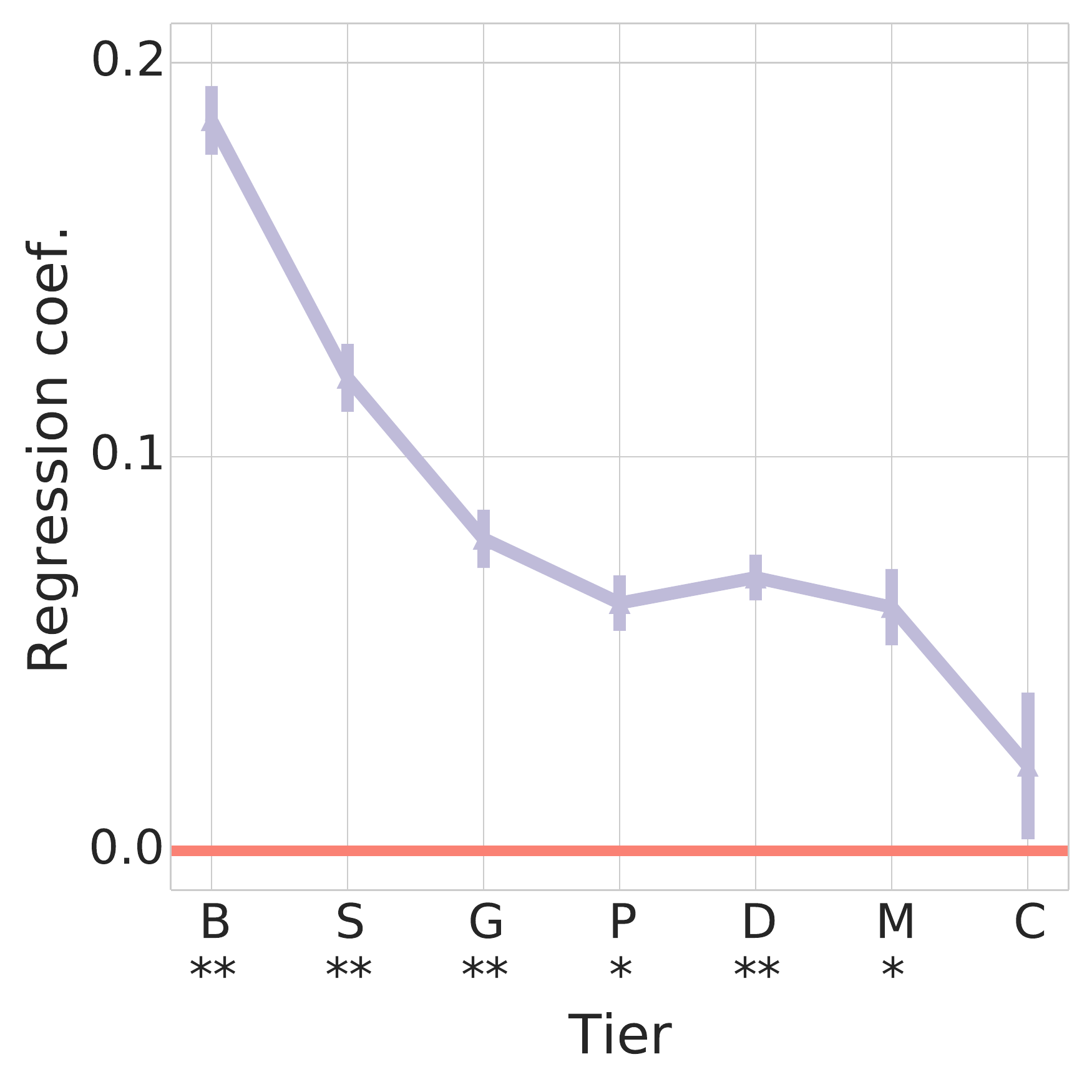}
  	\caption{Correlations coefficients between background diversity and team congruency by tier. Elite players are significantly less prone to proficient champion overlaps among team members.}\label{fig:reg_assigned}
 \endminipage\hfill
 \minipage{.6\columnwidth}
    \centering
    \includegraphics[width=\linewidth]{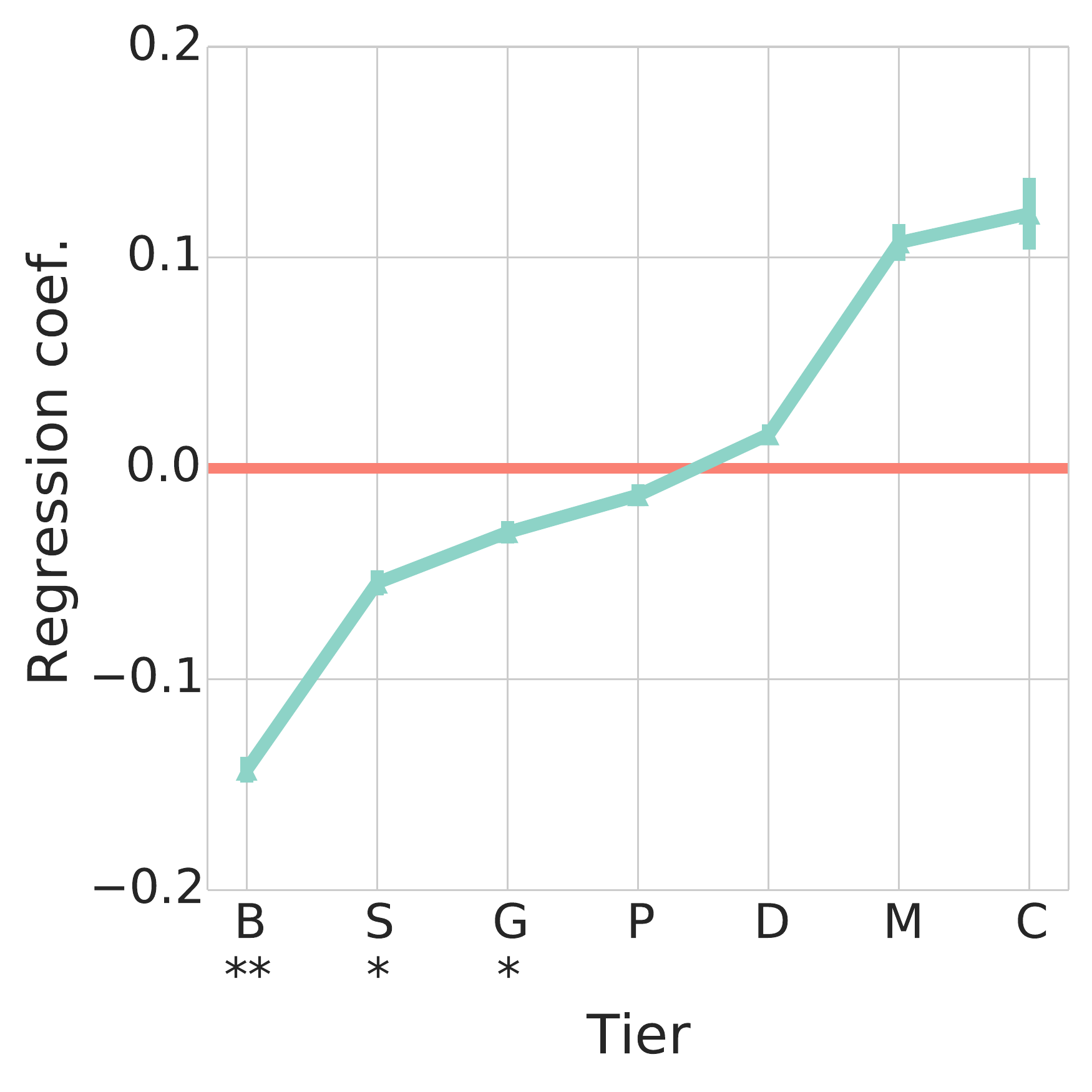}
    \caption{Correlations coefficients between team proficiency and team congruency by tier. Elite players maximize both proficiency and congruency significantly more than novice players. }\label{fig:reg_proficiency}
 \endminipage\hfill
 \minipage{.6\columnwidth}
    \centering
    \includegraphics[width=\linewidth]{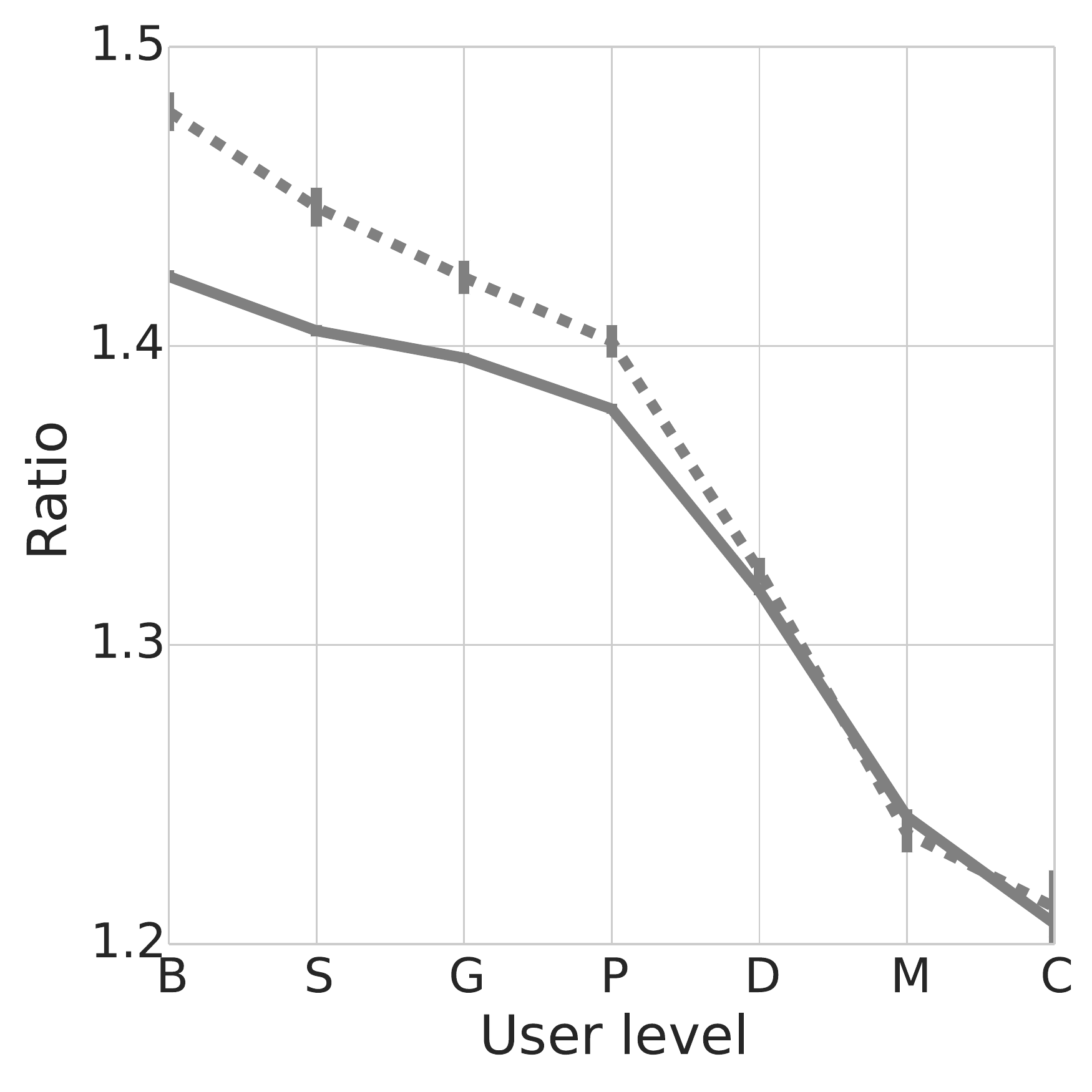}
    \caption{Ratio of the first pick to fifth pick proficiency by tier for all teams (solid) and teams in the bottom decile of background diversity (dotted). Novice players maximize proficiency more than elite players.}\label{fig:ratio}
 \endminipage\hfill
\end{figure*}

First we examine how the current team's congruency is influenced by the diversity of players' experience. If all members on a team have backgrounds using similar champions, then teams will have greater difficulty assembling proficient congruency. We estimate the correlation between a team's background diversity (Equation~\ref{eq:background_diversity}) and the team's congruency. Higher values indicate the observed team congruency is more influenced by players' background diversity. Figure~\ref{fig:reg_assigned} plots the regression coefficients by tier and the negative trend suggests elite teams' congruency is significantly less related to background diversity than novice players' congruency.

% First we examine how the current diversity of a team is influenced by the diversity of players' experience. If all members on a team have a history of using similar champions or fulfilling similar roles, then most players will end up playing in a non-proficient role. Controlling for team assignment features at the bottom of Table~\ref{tbl:features}, we estimate the correlation between team members' most-picked champions and the team diversity. Higher values indicate the observed team diversity is better-explained by players selecting their most-played champions. Figure~\ref{fig:reg_assigned} plots the regression coefficients for players in each tier. The negative trend across tiers suggests that teams of novice players are significantly most prone to selecting champions on the basis of who they play the most compared to expert players.

However, players do not have to sacrifice individual expertise to maximize team compatibility. Again using multiple regression to control for team assignment features, we estimate the correlation between average team proficiency and team diversity for players of different levels. Higher values indicate team members are playing their most proficient champions while also maximizing the diversity of the team. Figure~\ref{fig:reg_proficiency} plots the estimates by tier. There is a strong and significant positive trend towards more elite teams having a superior ability to simultaneously maximize their proficiency and congruency while novice players' team proficiency and congruency are significantly anti-correlated.%\brian{Check to make sure that when we discuss results about diversity, they're really about diversity and not congruency} \jooyeon{Figure 5,6 are the relationships with CONGRUENCY (we modified it from DIVERSITY)}

Finally, because the randomized pick order during ranked setup influences players' ability to select proficient champions, first pick players have a greater opportunity to select proficiency-maximizing champions than fifth pick players. Figure~\ref{fig:ratio} examines the ratio between the proficiency of the first pick's champion and the fifth pick's champion across tiers. We find a significant negative pattern suggesting that the difference between the proficiencies of elite teams is much smaller than for novice teams. This corroborates the findings from Figure~\ref{fig:proficiency}A that showed elite fifth pick players have significantly higher proficiency compared to novices.

The dashed line in Figure~\ref{fig:ratio} plots the pattern for players in the bottom decile of background diversity (Equation~\ref{eq:background_diversity}) in each tier. These are the teams composed of players having the most overlap between the champions they typically play and thus they face the hardest proficiency-congruency dilemmas. Elite players, even when they have substantial overlapping backgrounds and strong incentives to select their most proficient champion, minimize the proficiency gap between first pick and fifth pick more than novices.

\subsection{Qualitative findings}
During our interview sessions, we observed how elite players make conscious choices to negotiate the dilemma through communication during champion selection:

\begin{quote}\textit{
``I was the fifth pick in the last match... so I had to go to Bot... but I'm not good at either [bot role], I said please don't put me in Bot... and he swapped for me, it helped me a lot." (P4, Platinum 5)
}\end{quote}

Elite players use pick order strategically to help specialists lower in the order select higher-priority champions and then trade champions with them before the match starts, thereby optimizing proficiency and congruency. %Moreover, in extreme situation which is most of a team member's proficient roles are overlapped, the dilemma can't be resolved by swapping the role so that elite players give up before the match starts even though there is a penalty for dodging the game:

\begin{quote}\textit{
``I was afraid of picking a champion. What if there is another player who can only play Supporter besides me? It's a big trouble... if Supporter and AD-carry is picked earlier, I can't do anything..." (P3, Platinum 4) % if I played alone (not with other participants), I would dodge the game...
}\end{quote}

Another strategy elite players use to manage the dilemma is to intentionally select a non-proficient champion if there are unique opportunities for them to combine abilities with another player or counter a specific opponent selection:

\begin{quote}\textit{
``I think the combination (of bot duo) is crucial. I have played Lucian only for a few times, I was sorry for selecting that, but it's good combination with [the other champion]...so I selected it." (P3, Platinum 4)
}\end{quote}

%Also, P8 reported more directly for resolving the dilemma:

%\begin{quote}\textit{
%``If I'm proficient enough to play what is needed (for team composition), I'll take it (rather than choosing the most proficient champ)." (P8)
%}\end{quote}

On the other hand, while non-elite players are aware of the importance of optimizing proficiency and congruency, they fail to coordinate these picks within the allotted time:

\begin{quote}\textit{
``We should have talked to each other for the position assignment... Because it was not determined earlier, I had to hurry for picking my champ." (P19, Silver 5)
}\end{quote}

%Players make strategic use of the pick order and 
Elite players' communication and coordination during the selection process distinguishes them from novice players. Their strategic use of pick order to allocate champions to the most proficient players demonstrates one important way how they resolve the proficiency-congruency dilemma.

\section{Discussion}

Using observational data for 1.9 million \textit{League of Legends} matches, we proposed three hypotheses about the relationship between individual proficiency, role congruency, and team performance. To test these hypotheses, we operationalized and evaluated several quantitative constructs, employed a mixture of quantitative and qualitative methods, and found support for all three hypotheses. First, team performance is significantly and powerfully influenced by the similarity of a player's role with their most-experienced role (proficiency). Second, the extent to which the team is composed of complementary roles (congruency) has a  significant but moderate direct effect on team performance. Finally, expert users are better able to negotiate the \textit{proficiency-congruency dilemma} than novice users by jointly maximizing both when possible but sacrificing individual proficiency for team congruency. In the face of a classic social dilemma, elite player are more likely to cooperate (maximize team congruency) than to defect (maximize personal proficiency).

These findings build on an emerging stream of empirical scholarship that uses player behavior in games to understand more general social, psychological, and organizational phenomena~\cite{williams_mapping_2010}, while also complementing and challenging existing findings about team design, diversity, and task performance. Previous studies of teamwork that actually measure specific behavioral constructs of teamwork emphasize the importance of leadership, mutual monitoring, backup behavior, adaptability, and team orientation through shared mental models, communication, and trust~\cite{salas_is_2005}. Elements of all these processes are found in the champion selection team design process alone and the constructs we have developed may likewise find application in other settings. Based on this literature, we defined four distinct quantitative constructs that can be applied to team design processes in other online games as well as generalized to other organizational settings.  Excessive functional diversity was previously found to hinder team performance~\cite{bunderson_comparing_2002}, but we found a monotonic increase in team performance as a function of team congruence. We attribute this to the design of the game that rewards team diversity and penalizes role homophily~\cite{donaldson_mechanics_2015, johnson_all_2015, kou_playing_2014}.

\subsection{Implications for Design and Theory}
Team mental models, transactive memory systems, and collective intelligence are all forms of team cognition where the representation of knowledge about a team's task environment are shared, communicated, and developed~\cite{klimoski_team_1994}. The more models that team members share in common, particularly when teams are composed of unfamiliar individuals~\cite{zellmer-bruhn_when_2008}, the better they will be able to implicitly coordinate their work, especially in highly dynamic environments~\cite{harrison_time_2003, kozlowski_enhancing_2006, mathieu_team_2008}. Our results suggest expert players have significantly different schemas of the team design process --- independent from the functional and coordination skills necessary to be successful in the game itself --- that suggests team formation in \textit{LoL} can be used in general social processes like decision making, influence, and creativity~\cite{kerr_group_2004}.

Players of multiplayer online battle arenas, e-sports, and other team-based online games face difficult challenges for assembling a competitive team. Expert players have superior mental models of how in-game roles complement each other but novices have to develop these over time, which can be a frustrating experience that leads to customer churn. Knowledge about champion congruency could be extracted from character attributes, behavioral data, or subjective evaluations (all of which we did in this paper) to develop awareness cues for the champion selection process. These awareness cues could warn novice players against making particularly incongruent champion selections, show visualizations about team congruency, or include recommendations for which types of roles are ideal for the current team composition. The strong path dependencies on teams' ability to win based on initial team composition suggests the need to potentially re-calibrate champion abilities on-the-fly for players who do not recognize the importance of role proficiency and congruency~\cite{huang_mastering_2013, vicencio-moreira_now_2015}. Alternatively, market mechanisms could be used to incentivize players to learn and play other roles by making under-represented champions less expensive or over-represented champions more expensive.

\subsection{Limitations and future work}
While we used a mixture of qualitative and quantitative methods to explore and analyze the proficiency-congruency trade-off, future work could augment these efforts to understand the relationships between virtual team design and performance. One threat to validity is that the observed behaviors may fail either to persist or do not capture dynamics occurring over longer periods of time~\cite{mcgrath_groups_1993}. Players could complete survey instruments (\textit{e.g.},~\cite{wageman_team_2005}) which would allow researchers to map observed behavioral patterns back to well-validated constructs about teamwork and organizational behavior. The claims we made in this paper are only correlational, but the scale and granularity of that data provides substantial opportunities to use methods like propensity score matching~\cite{dehejia_propensity_2002} or use the the game's extensive randomization to identify natural experiments~\cite{rutter_proceeding_2007} to make stronger causal inferences.

MOBAs like \textit{LoL} have a global audience and players are clustered in geographically-bounded server regions such as North America, Western Europe, Latin America, and Korea.~\cite{riot_separate_severs} This provides a unique opportunity to conduct cross-cultural comparisons about organizational behavior around decision making, collaboration, and teamwork~\cite{gelfand_cross-cultural_2007}. While we reported findings aggregated across three regions, we observed interesting differences between regions that merit follow up analysis. Because MOBA players are almost exclusively male and users' genders are obscured, there is also substantial potential to explore how mixed-gender and non-male teams differ in team competencies~\cite{ratan_stand_2015}. 

This paper did not examine the role of social relationships and players' interaction histories on team performance. Prior research has emphasized the importance of social ties within multiplayer games for team performance~\cite{huang_functional_2013, mason_friends_2013, nardi_strangers_2006, xu_sociable_2011} and future work on team performance in MOBAs should explore the role of member familiarity and relationship strength on team performance~\cite{iosup_implicit_2014}. Constructs such as skill are also highly variable in terms of behaviors such as frequency, intensity, and variation in game style~\cite{huang_mastering_2013, reeves_experts_2009} and, following calls to examine the structure of behavioral sequences in socio-technical systems~\cite{keegan_sequence_2016}, the evolutions of players' champion selections could reveal interesting exploration-exploitation dynamics over the complex champion feature space~\cite{march_exploration_1991}.

%\section{Conclusion}
% computational social science, sports as models of data-rich team performance to inform 
% task congruence

%\section{Acknowledgments}
%\textit{Blinded for review.}
\newpage

% Balancing columns in a ref list is a bit of a pain because you
% either use a hack like flushend or balance, or manually insert
% a column break.  http://www.tex.ac.uk/cgi-bin/texfaq2html?label=balance
% multicols doesn't work because we're already in two-column mode,
% and flushend isn't awesome, so I choose balance.  See this
% for more info: http://cs.brown.edu/system/software/latex/doc/balance.pdf
%
% Note that in a perfect world balance wants to be in the first
% column of the last page.
%
% If balance doesn't work for you, you can remove that and
% hard-code a column break into the bbl file right before you
% submit:
%
% http://stackoverflow.com/questions/2149854/how-to-manually-equalize-columns-
% in-an-ieee-paper-if-using-bibtex
%
% Or, just remove \balance and give up on balancing the last page.
%
\balance{}

% REFERENCES FORMAT
% References must be the same font size as other body text.
\bibliographystyle{SIGCHI-Reference-Format}
\bibliography{proceedings}

\end{document}